\documentclass[12pt,preprint]{aastex}
\usepackage{apjfonts}
\usepackage{epsfig}
\usepackage{amsmath}
\begin{document}

\title{Testing Blandford-Znajek mechanism in black hole hyperaccretion flows for long-duration gamma-ray bursts}
\author{Mei Du$^{1}$, Shuang-Xi Yi$^{1}$, Tong Liu$^{2}$, Cui-Ying Song$^{3}$ and Wei Xie$^{4}$}
\affil{$^{1}$School of Physics and Physical Engineering, Qufu Normal University, Qufu 273165, China; yisx2015@qfnu.edu.cn\\
       $^{2}$Department of Astronomy, Xiamen University, Xiamen 361005, China\\
       $^{3}$Tsung-Dao Lee Institute, Shanghai Jiao Tong University, Shanghai 200240, China\\
       $^{4}$Guizhou Provincial Key Laboratory of Radio Astronomy and Data Processing, Guizhou Normal University, Guiyang 550001,China}

\begin{abstract}
Long-duration gamma-ray bursts (GRBs) are generally related to the core-collapse of massive stars. In the collapsar scenario, a rotating stellar-mass black hole (BH) surrounded by a hyperaccretion disk has been considered as one of the plausible candidates of GRB central engines. In this paper, we work on a sample including 146 long GRBs with significant jet break features in the multi-band afterglows. The jet opening angles can be then obtained by the jet break time. By asumming GRB jets powered by Blandford-Znajek (BZ) mechanism in the BH hyperaccretion system, we analyze the distributions of the long GRB luminosities and durations in the samples, and constrain the accretion rates for the different BH spins. As the results, we find that the BZ mechanism is so powerful making it possible to interpret the long GRB prompt emissions within the reasonably accretion rates.
\end{abstract}

\section{Introduction}

Gamma-ray bursts (GRBs) are supposed to be the extremely energetic transient events in the sky.
These erratic and high luminous transients are very different from the other astronomical phenomena.
Depending on a separation at 2 seconds of their $\rm \gamma-ray$ duration $T_{90}$, GRBs are sorted into two categories, long ($T_{90}>2 \rm ~s$) and short ($T_{90}<2 \rm ~s$) GRBs. Long GRBs are thought to be the core-collapse of a massive star (e.g., Woosley 1993; MacFadyen \& Woosley 1999), while short GRBs are supposed to the merger of two compact stars (e.g., Paczynski 1986; Eichler et al. 1989; Narayan et al. 1992).

Our understanding of GRBs has been greatly improved since 1997, and according to the fireball shell model, the broadband afterglow emissions are from the external shock emission as the interaction of an ultra-relativistic ejecta with the circumburst medium (e.g., M{\'e}sz{\'a}ros \& Rees 1997; Sari 1998; Wu et al. 2003; Zou et al. 2005; Yi et al. 2013). Lots of multi-wavelength afterglows have been collected, and some different emission features have been found in the multi-wavelength afterglows after the launch of \emph{Swift}, such as, the five components in the canonical X-ray lightcurves, including several power-law decay phases and the erratic X-ray flares (e.g., Zhang et al. 2006; Nousek et al. 2006; Burrows et al. 2005; Falcone et al. 2006, 2007; Liang et al. 2006; Chincarini et al. 2007, 2010; Abdo et al. 2011; Troja et al. 2015; Yi et al. 2015, 2016, 2017a; Si et al. 2018), and some smooth onset bump, sharp reverse shock emission or supernovae component in the optical lightcurves (e.g., Liang et al. 2010, 2013; Japelj et al. 2014; Gao et al. 2015; Yi et al. 2020; Zhou et al. 2020). However, compared with the GRB afterglow emission, the physical origin of the GRB prompt emission is still less clear, even though the widely discussed scenario is the internal shock model (e.g., Piran 2004; Zhang 2007; Kumar \& Zhang 2015). But in order to explain the prompt emissions, the central engine models of GRBs needs to be proposed.

Different types of GRBs should have different physical origin, but two leading central engines have been introduced: either a hyper-accreting stellar-mass black hole (BH, see e.g., Woosley 1993; Popham et al. 1999; Narayan et al. 2001; Liu et al. 2007; Lei et al. 2013; Xie et al. 2016; Yi et al. 2017b) or a rapidly rotating (the typical rotation period is about $1 \rm ms$), strongly magnetized (the dipole magnetic field strength is about $10^{15} \rm G$) neutron star (namely magnetar, e.g., Usov 1992; Dai \& Lu 1998a, b; Zhang \& M{\'e}sz{\'a}ros 2001; Metzger 2010; Rowlinson et al. 2013, 2014; L{\"u} \& Zhang 2014). Both types of central engines may be operating in GRBs, but according to the GRB observations, we found the majority of GRBs likely have a hyper-accreting BH central engine.

In the framework of the BH central engine, two mechanisms for powering GRB jet have been mentioned: neutrinos-antineutrinos annihilation above or below the BH accretion disks (e.g., Popham et al. 1999; Di Matteo et al. 2002; Gu et al. 2006; Kawanaka \& Mineshige 2007; Liu et al. 2007, 2016; Zalamea \& Beloborodov 2011; Xue et al. 2013; Song et al. 2016; Xie et al. 2016); and the Blandford-Znajek (BZ) mechanism (Blandford \& Znajek 1977) tapping the spin energy of the BH through large scale magnetic fields (e.g., Lee et al. 2000a, 2000b; McKinney 2005; Nagataki 2011; Lei
et al. 2013, 2017; Wu et al. 2013; Liu et al. 2015). However, it is unsure which one is the primary mechanism to launch the GRB jet.
Fan \& Wei (2011), Liu et al. (2015) and Song et al. (2016) have investigated whether the neutrino-dominated accretion flows (NDAFs) could power GRBs (including some short and long GRBs). They pointed out the disk mass of the selected
GRBs mainly depends on the output energy of the NDAFs, jet opening angle, and characteristics
of central BH. As a result, they found that the NDAF model works well in lots of GRBs, but still have difficulty in explaining some energetic bursts. More details about the NDAFs can be seen the review work for Liu et al. (2017).

However, Yi et al. (2017b) compared the observational results with neutrino annihilation and BZ mechanisms, they found the BZ mechanism could generally account for the selected GRB data, which also can be seen in Lei et al. (2013). The result is further confirmed by Xie et al. (2017).
Tchekhovskoy et al. (2011) simulated a large amount of magnetic flux entering the BH, and the excess magnetic flux was blocked to form a magnetoresistive disk and produced outflows, this process was called magnetoresistive accretion. They found for the BH spin value $a_\bullet$ = 0.99, the energy flow into the BH ergosphere was less than that out, thus confirming that the BH would extract the rotating energy of the BH through the BZ mechanism.
Therefore, we tentatively assume that the jet is driven by the BZ mechanism in this work, and in order to test the ability of the BZ mechanism model, long GRBs with jet break features are needed.
This work is organized as follows. In Section 2, we present the BH central engine model. Section 3, some GRBs with jet breaks are listed to test the BZ mechanism model, and the constraint results for the accretion rate with a BZ jet are shown in Section 4.
Discussion and Conclusions are presented in Section 5.

\section{The BH central engine model}
After a massive star collapses, a stellar-mass BH surrounded by an accretion disk might born in the center. The BH accretion system could produce relativistic jet by neutrino annihilation or BZ mechanism. It has been proved that the strongly magnetized BZ mechanism is more effective than the non-magnetized neutrino annihilation processes to power GRB jet for the same BH spin parameter and accretion rate (e.g., Liu et al. 2015, 2018). In the collapsar model, when the jets break out from the envelope, the accretion rate decreases at $\lesssim 0.001~M_\odot~\rm s^{-1}$(e.g., Liu et al. 2018), which approaches the lower limit of the ignition accretion rate of NDAFs (e.g., Chen \& Beloborodov 2007; Zalamea \& Beloborodov 2011).

Thus we reasonably assume that the BZ process as the GRB jet production mechanism in this work, which is connected with its accretion rate $\dot{M}_{\rm in}$ (Blandford \&
Znajek 1977; Lee et al. 2000a, b):
\begin{equation}
\dot{E}_{\rm BZ}=1.7\times10^{20}a_\bullet^{2}m^{2}B_{\rm in, G}^{2}F(a_\bullet)~{\rm erg~s^{-1}},
\end{equation}
where $B_{\rm in, G}=B_{\rm in}/1~\rm G$ is the magnetic field strength threading the BH
horizon, $m=M_{\rm BH}/M_{\odot}$, the spin parameter $a_\bullet$ and
\begin{equation}
F(a_\bullet)=[(1+q^{2})/q^{2}][(q+1/q)\arctan(q)-1]
\end{equation}
is a spin-dependent dimensionless parameter, and $q=a_\bullet/(1+\sqrt{1-a_\bullet^{2}})$. With the balance between the ram pressure of the innermost part of the disk $P_{\rm in}$ and the magnetic pressure on the BH horizon, one has
\begin{equation}
\frac{B_{\rm in}^{2}}{8\pi}=P_{\rm in} \sim \rho_{\rm in} c^{2}\sim\frac{\dot{M}_{\rm in}c}{4\pi r_{\rm H}^{2}},
\end{equation}
where $r_{\rm H}=(1+\sqrt{1-a_\bullet^{2}})r_{\rm g}$ means the radius of the BH horizon, $\dot{M}_{\rm in}$ and $\rho_{\rm in}$ represents the net accretion rate and density at the inner boundary of the disk, respectively.

We can then estimate the magnetic field strength threading the BH
horizon as
\begin{equation}
B_{\rm in}\simeq 7.4\times10^{16}  \dot{m}_{\rm in}^{1/2} m^{-1}(1+\sqrt{1-a_\bullet^{2}})^{-1}~ \rm G,
\end{equation}
where $\dot{m}_{\rm in}=\dot{M}_{\rm in}/(M_{\odot}~\rm s^{-1})$. Taking above equation into Equation (1), the BZ jet luminosity can be rewritten as
\begin{equation}
\dot{E}_{\rm BZ}=9.3\times10^{53}a_\bullet^{2}\dot{m}_{\rm in} F(a_\bullet)(1+\sqrt{1-a_\bullet^{2}})^{-2}~{\rm erg~s^{-1}}.
\label{eq:BZ}
\end{equation}
This indicates that the BZ power does not depend on the mass of the BH (Lei et al. 2011).
On the other hand, the observed mean GRB jet luminosity $L_{\rm \gamma}$ is generally a fraction of the total BZ power of the GRB averaged in the whole accretion process, i.e.,
\begin{equation}
L_{\rm \gamma}=\eta \dot{E}_{\rm BZ},
\end{equation}
where $\eta$ is radiant efficiency.

The hyperaccretion inflow with a BZ jet may result in the violent evolution of the BH, and give some physical information of progenitor stars of long GRBs. Therefore, it makes more sense to take more GRB samples for constraining the accretion rate with a BZ jet.
We can get the limited range of the accretion rate when using the catalogue of long GRBs at different distances with vary durations.

\section{GRB Sample}
Large GRB sample and multi-wavelength afterglow observations may provide us a good opportunity to research
the properties of the GRB progenitors and  central engine models.
In order to test the BH central engine, the selected GRBs with the beaming corrected energies/luminosities are required.
Some temporal jet-break like features have been appeared in multi-wavelength light curves prove that the GRB outflows may be
collimated, and the temporal transition can be interpreted as a jet break which connects a normal decay phase ($\sim - 1$) to a steeper phase segment ($\sim - 2$) (e.g., Rhoads 1999; Sari et al. 1999; Frail et al. 2001; Wu et al. 2004; Liang et al. 2008; Racusin et al. 2009; Wang et al. 2018). More details about jet breaks can be seen in Zhao et al. (2020). They have collected the large sample with significant jet breaks appeared in the multi-wavelength afterglow light curves.

Under the criteria of the light curves transform from a normal decay phase to a steeper phase segment from multi-wavelength afterglow light curves, we extensively search for the GRBs with significant jet breaks, and 146 long GRB candidates with such jet breaks are collected. The majority of GRB data are taken from Zhao et al. (2020), also including three ultra-long GRBs 060218, 090417B and 111209A. Most of the jet breaks are identified by X-ray light curves, the others are calculated from optical and radio afterglows. The relevant information of the GRB sample (such as, the redshift $z$, the duration $T_{90}$, the isotropic energy $E_{\rm \gamma, iso}$, the jet break time $T_{\rm jet}$ and the references) is presented in Table 1. The values of $T_{90}$ of GRB sample vary from a few seconds
to about 25000 seconds, and the redshifts of GRBs have a wide range.
For jet-break GRBs, we derive the jet opening angles in a homogeneous interstellar medium (e.g., Rhoads 1999; Sari et al. 1999; Frail et al. 2001; Yi et al. 2015; 2017b) with,
\begin{equation}
\theta_{\rm jet}=0.076 \,\,{\rm rad}\,\,\left(\frac{T_{\rm jet}}{1\ \rm
day}\right)^{3/8}\left(\frac{1+z}{2}\right)^{-3/8}
E_{\rm \gamma,iso,53}^{-1/8}\left(\frac{\eta_{\gamma}}{0.2}\right)^{1/8}\left(\frac{n}{1\ \rm
cm^{-3}}\right)^{1/8},
\end{equation}
where the efficiency of the fireball in converting the energy
in the ejecta into $\gamma$-rays $\eta_{\gamma}=0.2$ and the number density of ambient medium $n=1$ cm$^{\rm-3}$ are general adopted. It should be noted that there are some uncertainties in the process of estimating jet opening angle, for instance, the unknown ambient density profile, efficiency parameter and coefficient. Following Frail et al. (2001), the coefficient we used here is 0.076. We also used the coefficient value 0.084 (see e.g., Zhang 2018) to calculate the accretion rate, the difference in result is minimal. Take GRB 970508 as an example, the first GRB in our work, the difference of the accretion rate is only 0.00094 when using different coefficient. With the half opening angle derived by selected sample, we can perform a measurement of the true luminosities from isotropic values of GRBs, i.e., $L_{\rm \gamma}=(1-\rm cos\theta_{jet})L_{\rm \gamma,iso}=L_{\rm \gamma,iso}\theta_{jet}^{2}/2$, where the mean isotropic luminosity of GRBs $L_{\rm \gamma,iso}=(1+z)E_{\rm \gamma,iso}/T_{90}$. Thus, the accretion rate of BH with the BZ mechanism can be estimated with the above equations and the observational data.

\section{Results and Discussion}
In our analysis, the observational GRB data are collected to derive the half jet opening angle and the collimation-corrected jet luminosity $L_{\gamma}$ for those long GRBs. Figure 1 shows the histogram distributions of the selected GRB parameters, the durations in the rest frame $T_{90,\rm i}=T_{90}/(1+z)$ of the selected GRB sample mainly from about 10~s to 1000~s, and three ultra-long GRBs (060218, 090417B and 111209A) are also collected, of which durations are longer than 2000~s. The distributions of
the GRB luminosity peak at few times of $10^{51}~\rm erg~ s^{-1}$ for the isotropic ones and a few times of $10^{48}~ \rm erg~s^{-1}$ for the collimation-corrected GRBs.

BHs are mysterious and energetic compact objects, in order to test the BZ process as the GRB jet production mechanism in this work, two essential values of BHs are needed, i.e., the accretion rate and spin. On the other words, once the accretion rate and BH spin parameter are provided,
we then could calculate the BZ jet power using Equation (5). The BH spin value is an important factor for the occurrence of GRBs, considering the BZ mechanism extract the spin energy of the BH through a magnetic field, the BZ jet power of GRBs are mainly dependent on the BH spin, not accretion rate. Moreover, the BZ power is larger than neutrino annihilation power for about two orders when using the same BH spin value and accretion rate (e.g., Kawanaka et al. 2013; Liu et al. 2015; Lei et al. 2017). Therefore, the larger BH spin parameter and smaller BH accretion rate in BZ process may appear when compared with neutrino annihilation process. So we apply the large spin parameters (0.5, 0.7, 0.87 and 0.998) and the radiant efficiency $\eta=0.2$ to our investigated GRBs. The constraint results on the accretion rate with different spin values of BHs in GRBs are shown in Table 2. Figure 2 presents the distributions of the accretion rates for different long GRBs under the BZ mechanism with the different typical spin parameters 0.5, 0.7, 0.87 and 0.998, respectively. And Figure 3 shows the accreted masses for the different typical spin parameters with our selected GRBs.

For the spin parameter $a_\bullet=0.5$, the fitted accretion rates distribute in a relatively wide range from $\sim 10^{-6} ~M_\odot\ \rm s^{-1}$ to $\sim 0.1~ M_\odot\ \rm s^{-1}$, by multiplying the samples' durations we get the accreted masses ranging from $\sim 10^{-5}~ M_\odot$ to $\sim 3 ~M_\odot$. These results are totally reasonable, considering the simulation results of the accretion disk mass $\sim5~ M_\odot$ (e.g., MacFadyen \& Woosley 1999; Popham et al. 1999; Zhang et al. 2003). The significant difference of the accretion rates among the samples might indicate the diversity of the magnetic field of the progenitor stars. Expectedly, the accretion rate is negatively correlated with the BH spin, the central values of accretion rate are $5.25 \times 10^{-4} M_\odot\ \rm s^{-1}$, $1.95 \times 10^{-4} ~M_\odot\ \rm s^{-1}$, $8.24 \times 10^{-5} ~M_\odot\ \rm s^{-1}$ and $2.57 \times 10^{-5} ~M_\odot\ \rm s^{-1}$ corresponding to $a_\bullet = 0.5$, $0.7$, $0.87$ and $0.998$, respectively. Then the central values of the accreted masses for each GRB are $1.28 \times 10^{-2} M_\odot\ $, $4.79 \times 10^{-3} ~M_\odot\ $, $2.34 \times 10^{-3} ~M_\odot\ $ and $6.92 \times 10^{-4} ~M_\odot\ $ for different spin parameters. It's obvious that the results sensationally depends on the spin parameter. Unfortunately, the BH spins are hard to constrain. By methods respectively based on disk continuum and reflection spectra, researchers have access to measure the BH spins in stellar-mass X-ray binaries. It's unlikely for the BH in a binary to accrete enough mass from the companion to change its mass and spin significantly, which might make it be a faithful proxy for the mass and spin of the newborn BH in a supernova/long-GRB. The results show that these BHs probably possess relatively high spins (Miller \& Miller 2015 and references therein). It's worth mentioning that for some energetic GRBs (e.g. GRBs 990123, 021004, 050820A, 060124, 060210, 070125, 090323, 090926A, and 130427A), it's difficult to fit them through NDAFs model due to its relatively lower output power (e.g., Song et al. 2016), leaving the BZ mechanism as the more possible underlying central engine. In addition, the required accretion mass $M_\mathrm{acc}$ for several special GRBs under a moderate spin (i.e., $a_\bullet = 0.5$) come up to very high values, such as $2.32 ~M_\odot$ for GRB 090323 and $3.47~ M_\odot$ for GRB 160625B. Such a huge amount of accreted mass will significantly increase the mass and spin of the BH and consequently influence our results. It will be meaningful to consider the evolution of the central BH for the more-energetic GRBs in the future. On the other side, some GRBs have relatively low accretion rates corresponding to their low luminosities(e.g., GRBs 060218 and 111209A). It's worth noting that the low-luminosity and ultra-long GRB 111209A is associated with a super luminous supernova (SLSN) SN 2011kl (Greiner et al. 2015).

If the radiation cooling is hard to balance the viscous heating in the vertical direction of the optically thick or thin disks, there may exist strong outflows from the disks (e.g. Gu 2015). Moreover, the photons might exert the radiation pressure upon the disk materials to blow them away in the optically thick disks, as well as the magnetic pressure in all type of the disks (e.g., Song et al. 2018). In whichever cases, the disk outflows cause the net inflow rates decrease, thus even though the BZ jets are dominated, the required accretion rates should be larger than the above results.

Although the BZ mechanism is an effective process to power GRBs, this does not mean that all GRBs have to be powered by the BZ mechanism. It has been suggested that the neutrino annihilation process is still a suitable candidate for generating GRBs, and the NDAFs can be applied to most of GRBs except for some cases with extremely high luminosity (e.g., Song et al. 2015). Neutrino annihilation and BZ mechanism might coexist in NDAF with magnetic coupling (MC), and compete with each other. Variations in the proportions of the BZ and MC components lead to changes in the dominance of the BZ power and the neutrino annihilation luminosity (e.g., Lei et al. 2009; Song et al. 2020).

In principle, a quickly rotating magnetar is another popular model for the central engines of GRBs, with the features of rapidly spinning and strongly magnetized. The spin-down or differential rotation of magnetars has been used to interpret some GRBs, especially for those GRBs with plateau phases in X-ray afterglows. The total rotational energy of a magnetar can be estimated with the value of $\sim 2 \times 10^{52}$ erg, making it possible to power most of GRBs and reconcile their diverse behaviors. However, also for this reason, some very bright GRBs (with the isotropic energy larger than $ 2 \times 10^{52}$ erg) should not satisfy a magnetar central engine with the assumption of isotropic energy emit from a quickly rotating magnetar. If the magnetar wind can be collimated, as shown in some numerical simulations of Bucciantini et al. (2009), a magnetar engine can power a GRB with isotropic energy much greater than $2 \times 10^{52}$ erg as long as the beaming-corrected energy falls below such a value (L{\"u} \& Zhang 2014; Li et al. 2018). Moreover, according to L{\"u} et al. (2015), the magnetar engines in short GRBs, on the other hand, can be considered as isotropic. Such energy limitation is not appropriate for the BH hyperacctrion models, therefore, most of GRBs likely satisfy a BH accretion central engine. Some GRBs with X-ray flares or plateau phase have also been collected in our sample, such as, GRBs 060714 (Krimm et al. 2007a) and 060729 (Grupe et al. 2007), and the central engine of those GRBs may be a magnetar. Here our purpose is just to derive and check the BH parameters under the BZ mechanism by using the large GRB sample with significant jet break features in the multi-wavelength afterglows. Therefore, the type of GRB
central engines should be still further studied even if we adopt the BZ mechanism to all of the selected GRBs.

\section{Conclusions}
Long-duration GRBs are thought to be the core-collapse of a massive star, and the leading model about GRB central engine relates to a stellar-mass BH surrounded by a hyper-accreting disk. Two mechanisms for powering GRB jet are hence mentioned in the framework of the BH central engine: neutrinos annihilation and BZ mechanisms, although it is unsure which one is the primary mechanism to launch the GRB jet.
Considering the strongly magnetized BZ mechanism is more effective processes to power GRB jet, we here mainly assume that the BZ process as the jet production mechanism.
In this work, we collect 146 long GRBs sample with significant jet break features in the multi-wavelength afterglows. The jet opening angles of those GRBs are then determined by the jet break time appeared in the lightcurves. We then derive and check the BH parameters under the BZ mechanism, and find that our selected GRB sample are consistent with the expectations of the BH model. We find the accretion rate is negatively correlated with the BH spin and conversion efficiency. We constrain accretion rates of BH with BZ process in the reasonable range by using the selected GRB sample.
As the results, we find that the BZ mechanism is so powerful making it possible to interpret the LGRB prompt emissions within the reasonably accretion rates. Especially for some energetic GRBs,
it's difficult to fit them through the neutrino mechanism due to its relatively lower output power
(e.g., Song et al. 2016), leaving the BZ mechanism as the more possible underlying central engine.

Moreover, how to identify a GRB central engine activity duration time for a compact star is still an open question, since some peculiar phenomena related to the activities of central engine have been observed, e.g., plateaus and X-ray flares. It likely indicates that the activity timescale of the GRB central engine may be much longer than the observed $\gamma$-ray $T_{90}$ duration of prompt emission phase. And thus, the more accreted masses are then required than our discussed in this work, if a BH hyperaccretion system is taken as the GRB central engine.
However, as we known, the jet power of the most GRBs decrease with time, and the majority energy is still released during
$T_{90}$. In any case, the collapsar signature may need further investigation with more observed data and the special phenomena related to the activities of GRB central engines.

\section{Acknowledgments}
We thank the anonymous referee for constructive and helpful comments. 
This work is supported by the National Natural Science Foundation of China (Grant Nos. 11703015, U2038106, 11822304 and 11847102),
the Natural Science Foundation of Shandong Province (Grant No. ZR2017BA006), and the Youth Innovations and Talents Project of Shandong Provincial Colleges and Universities (Grant No. 201909118).

\clearpage
\begin{figure*}
\includegraphics[angle=0,scale=0.30]{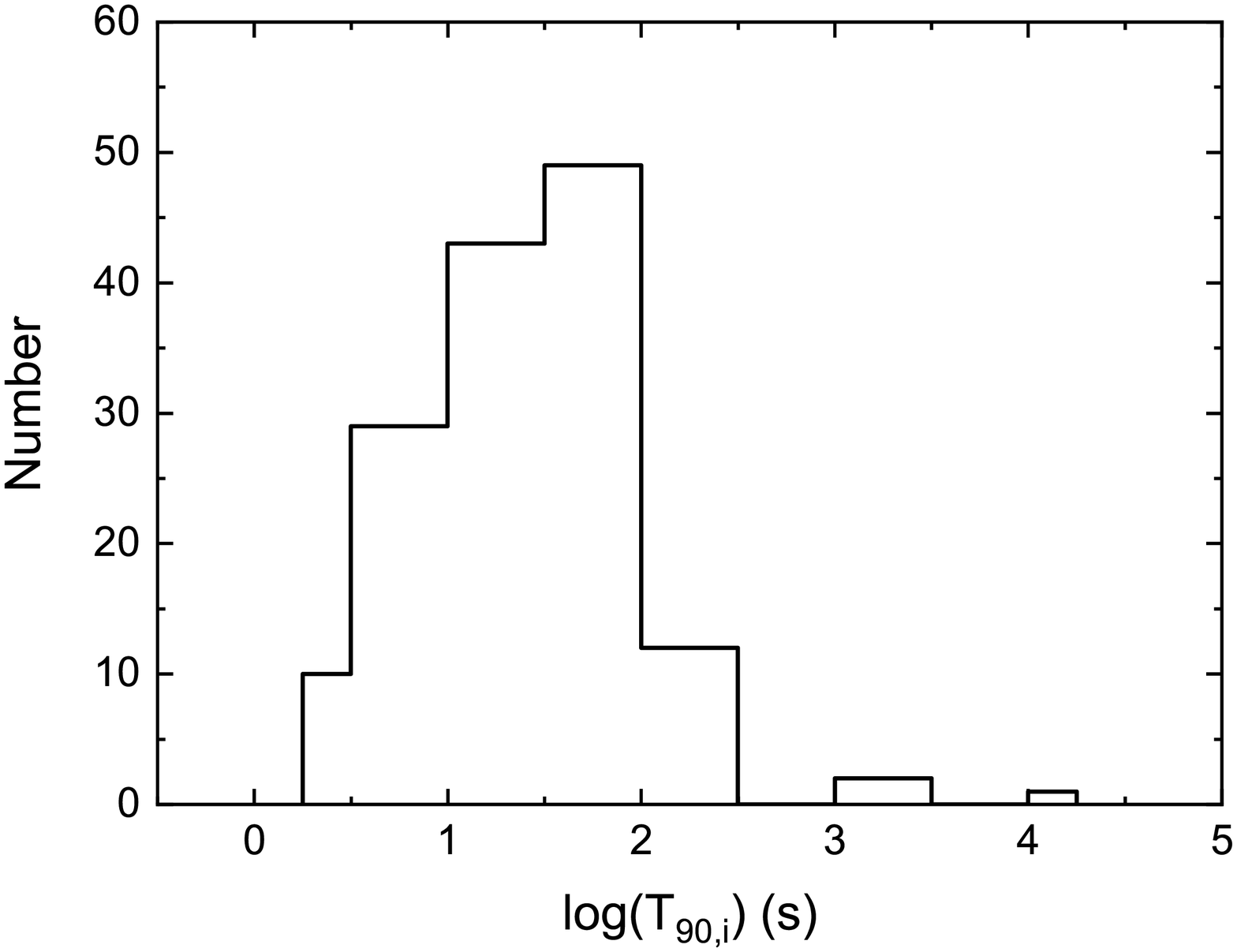}
\includegraphics[angle=0,scale=0.30]{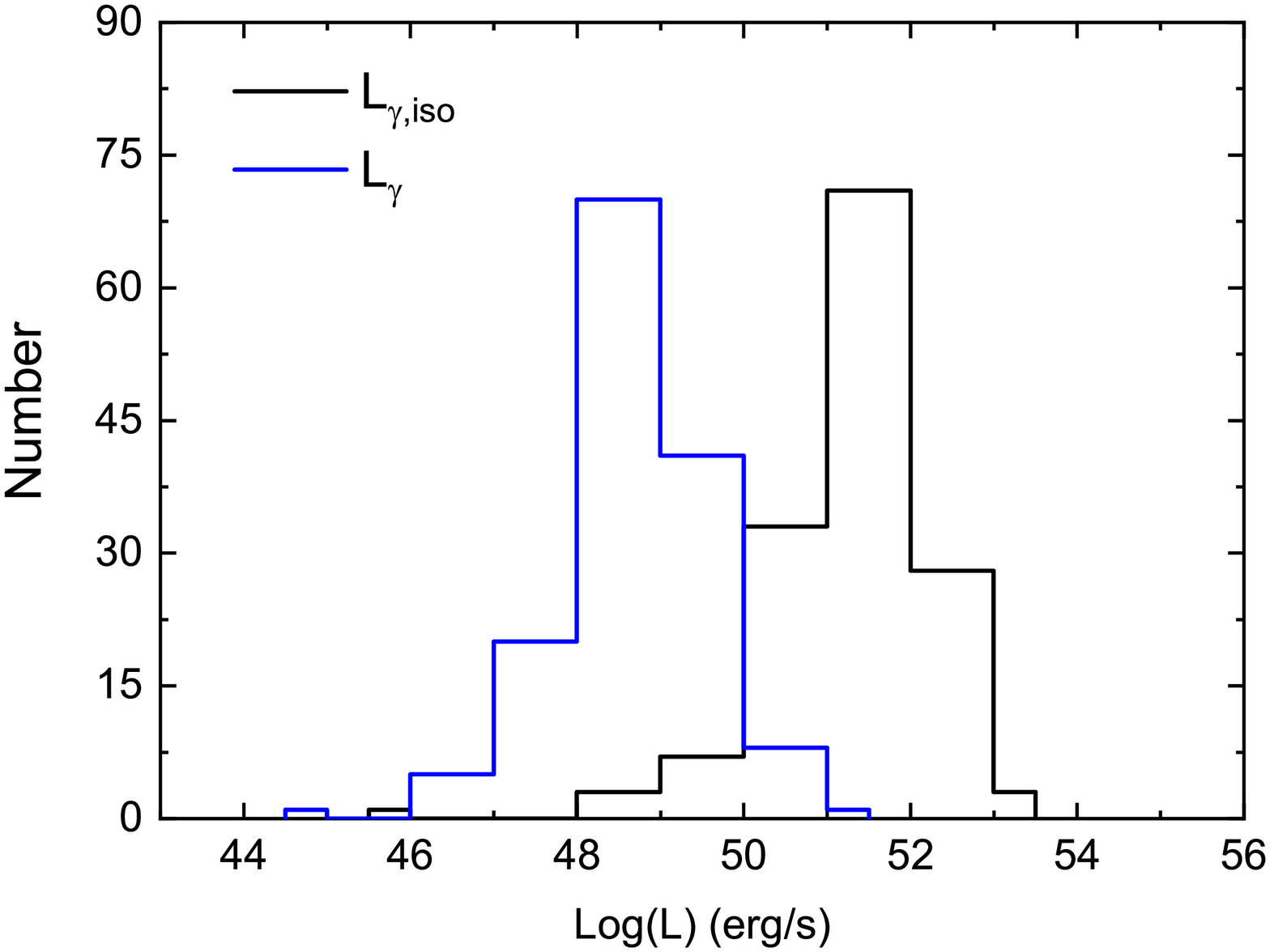}
\caption{Distributions of the duration and luminosity for collected GRBs.}
\end{figure*}

\begin{figure*}
\includegraphics[angle=0,scale=0.30]{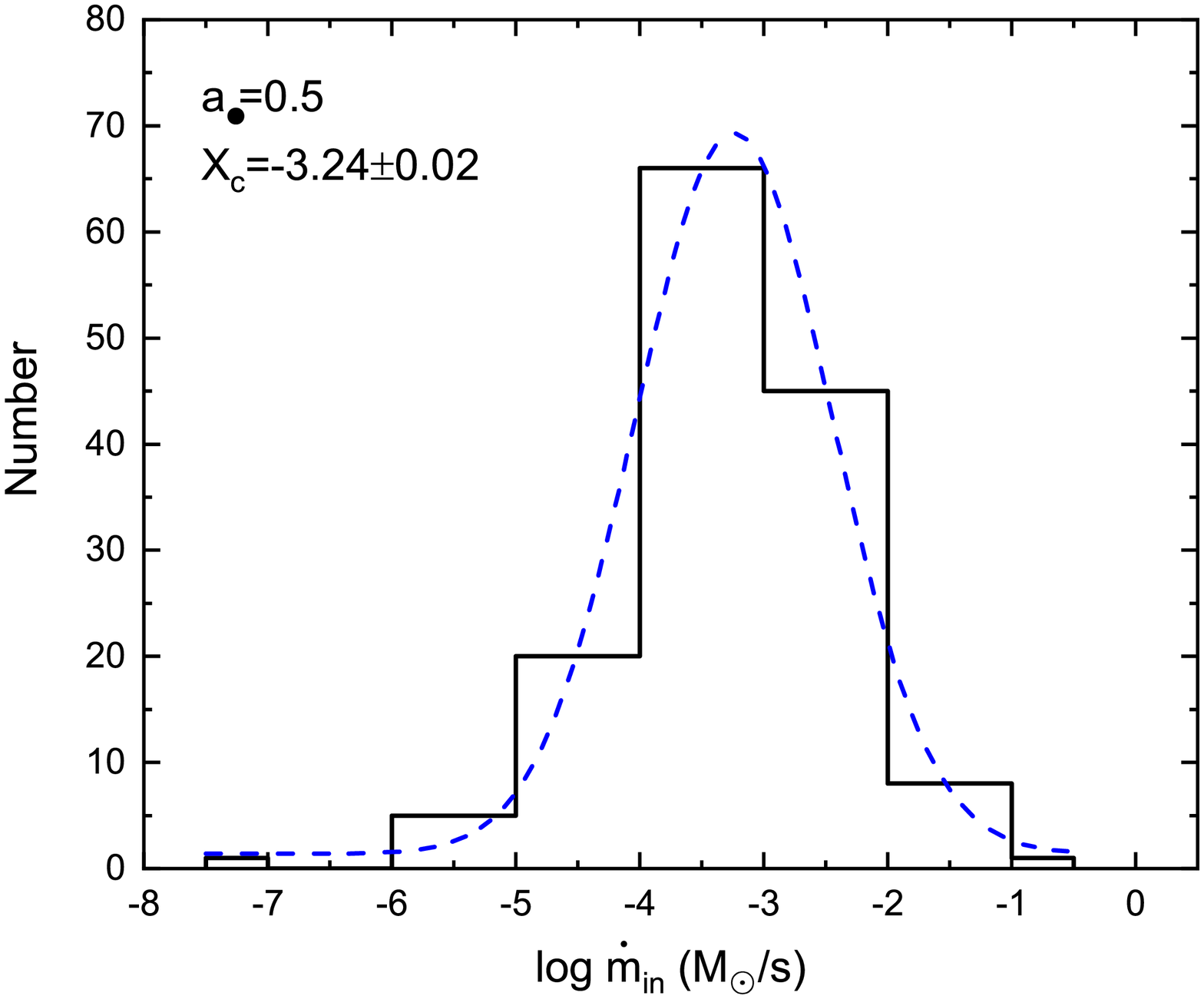}
\includegraphics[angle=0,scale=0.30]{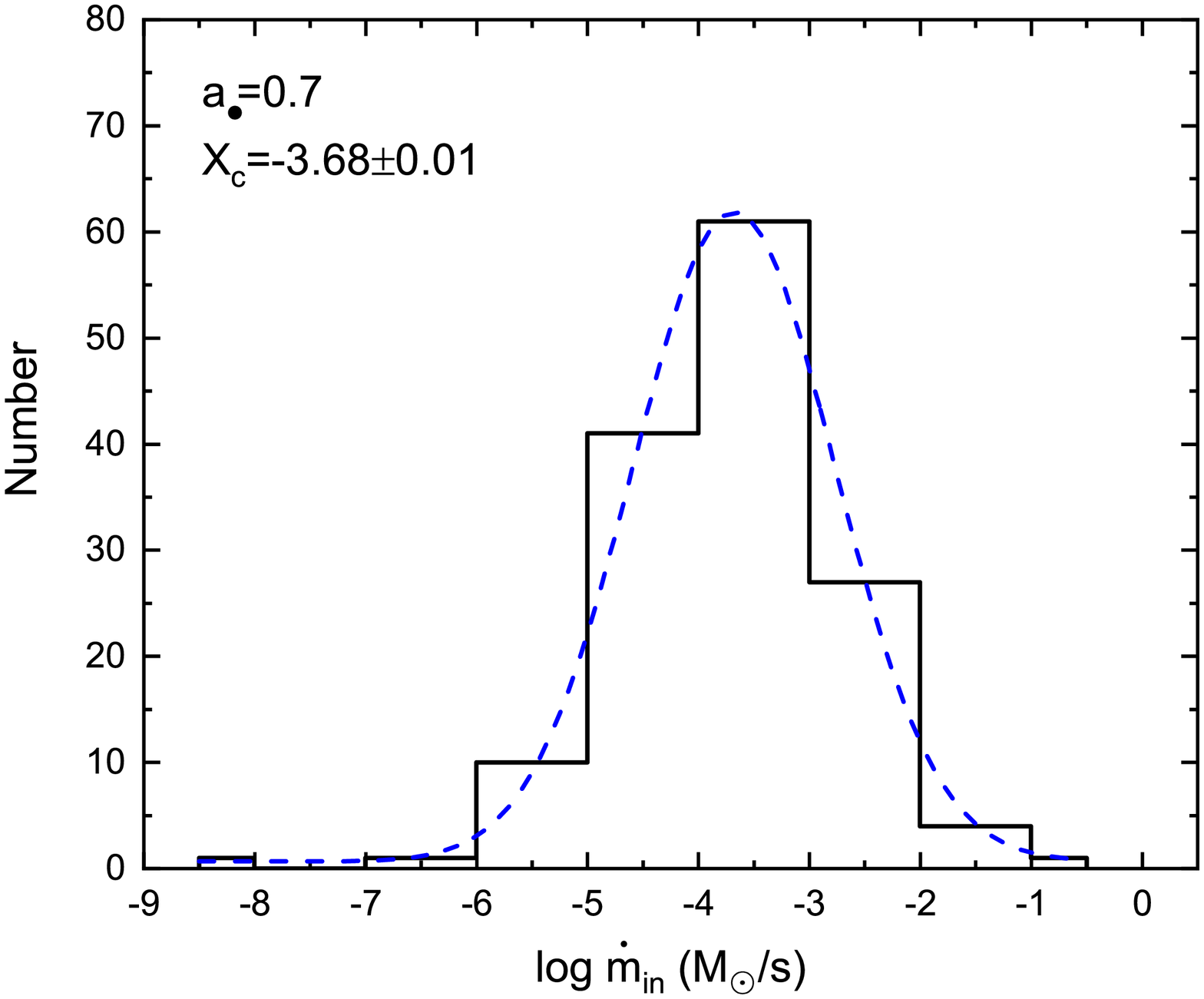}
\includegraphics[angle=0,scale=0.30]{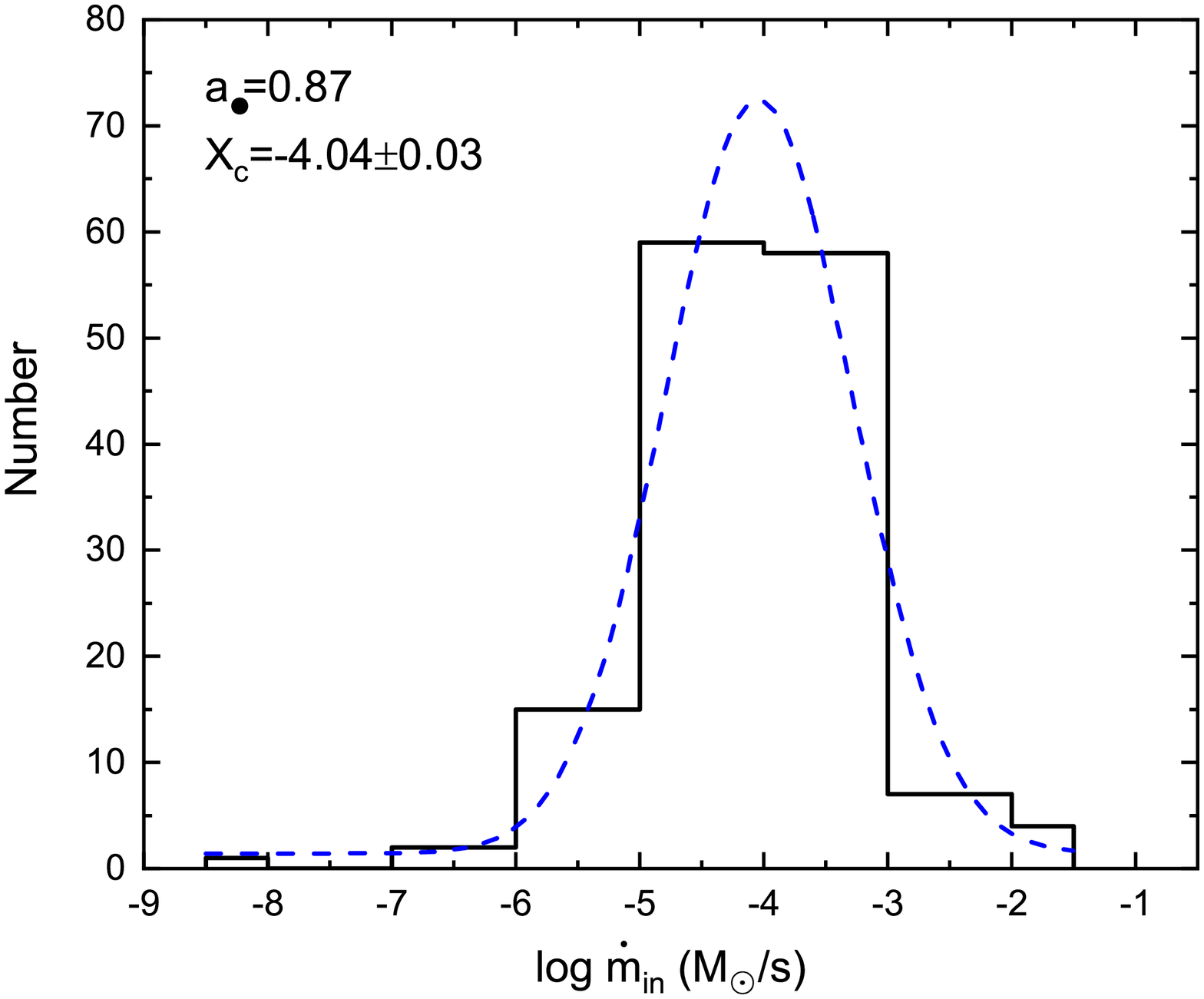}
\includegraphics[angle=0,scale=0.30]{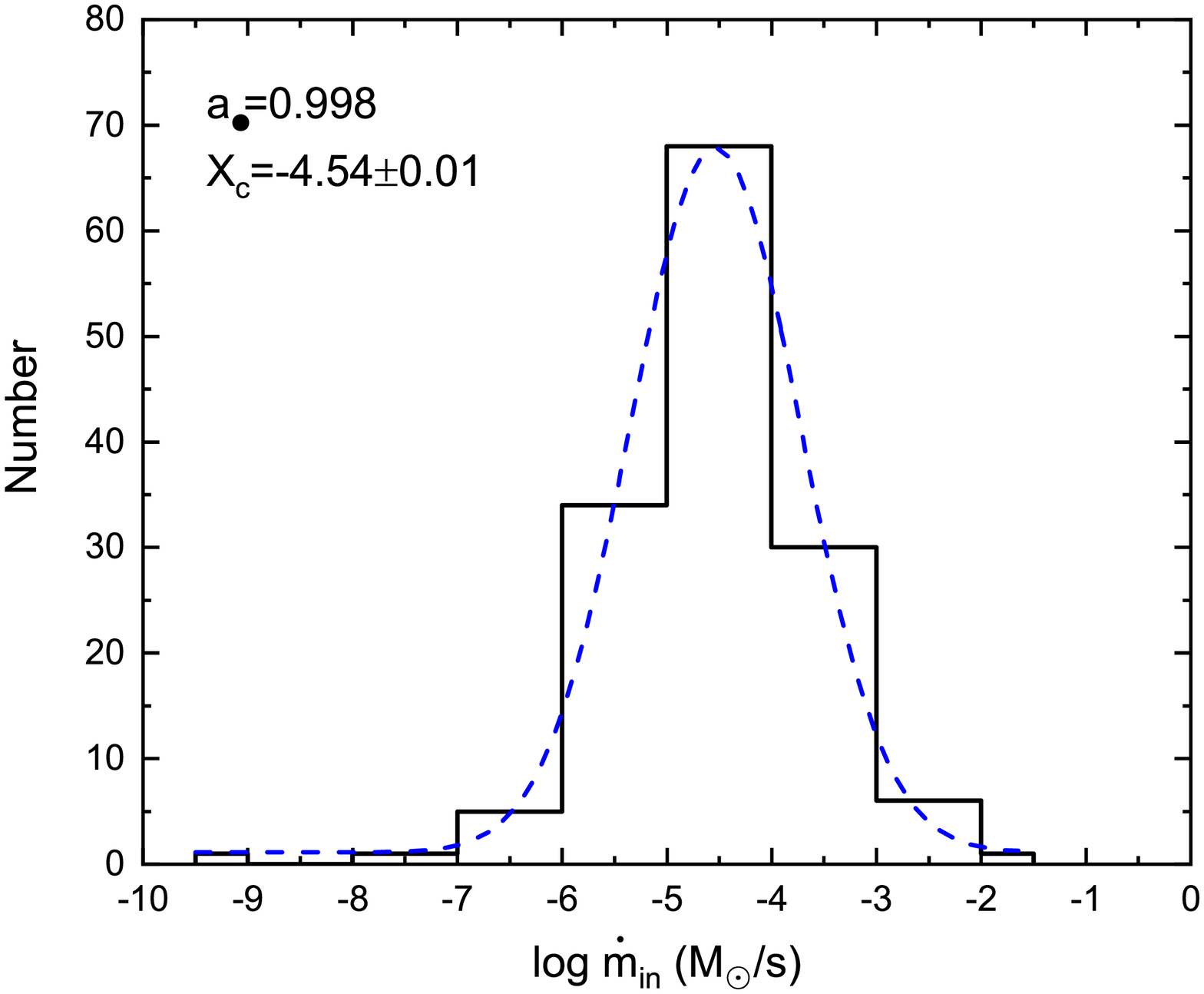}
\caption{Distributions of the accretion rate for different typical spins $a_\bullet$ with our selected GRB sample. The blue dashed line is the Guass fitting line.}
\end{figure*}

\begin{figure*}
\includegraphics[angle=0,scale=0.30]{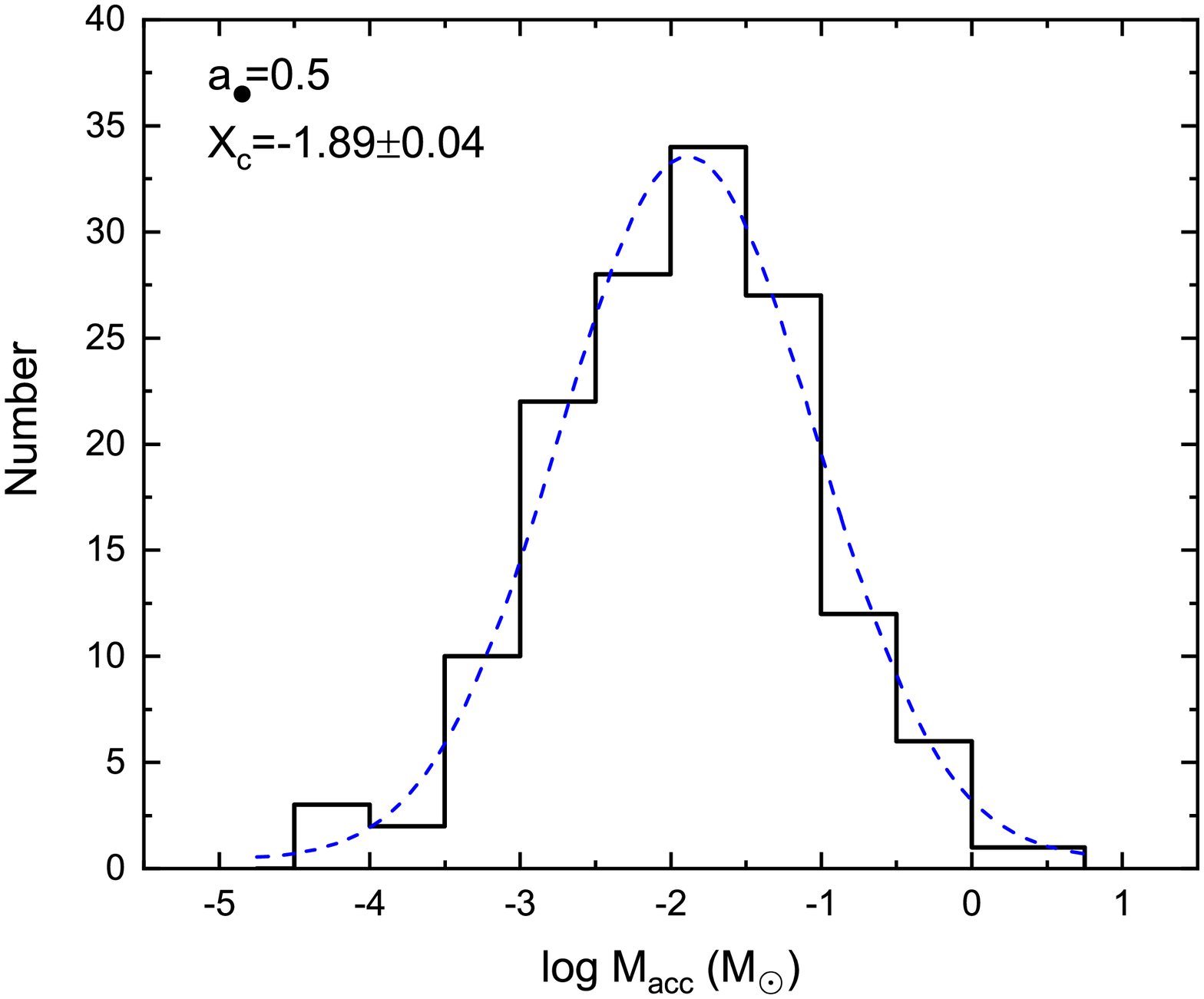}
\includegraphics[angle=0,scale=0.30]{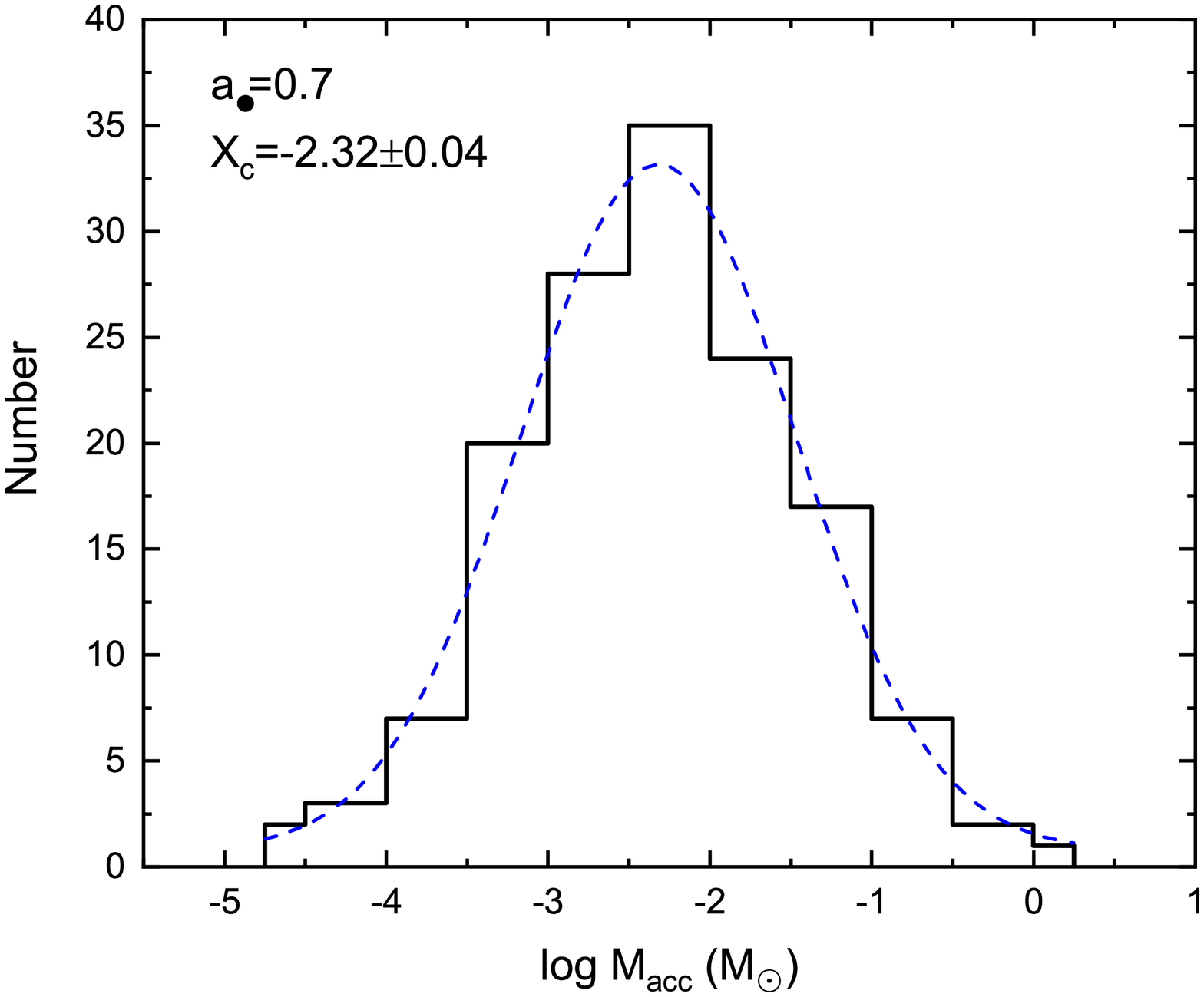}
\includegraphics[angle=0,scale=0.30]{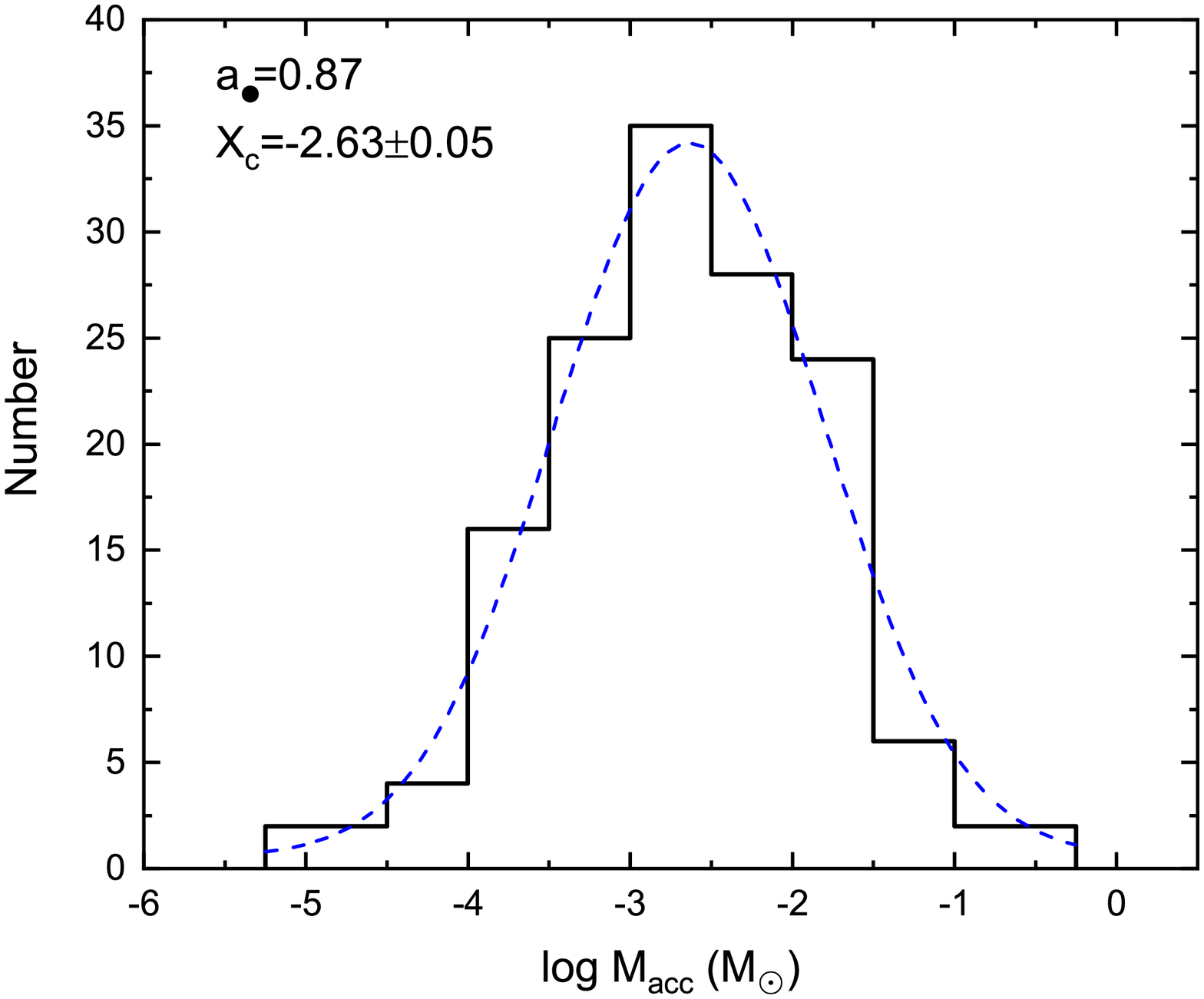}
\includegraphics[angle=0,scale=0.30]{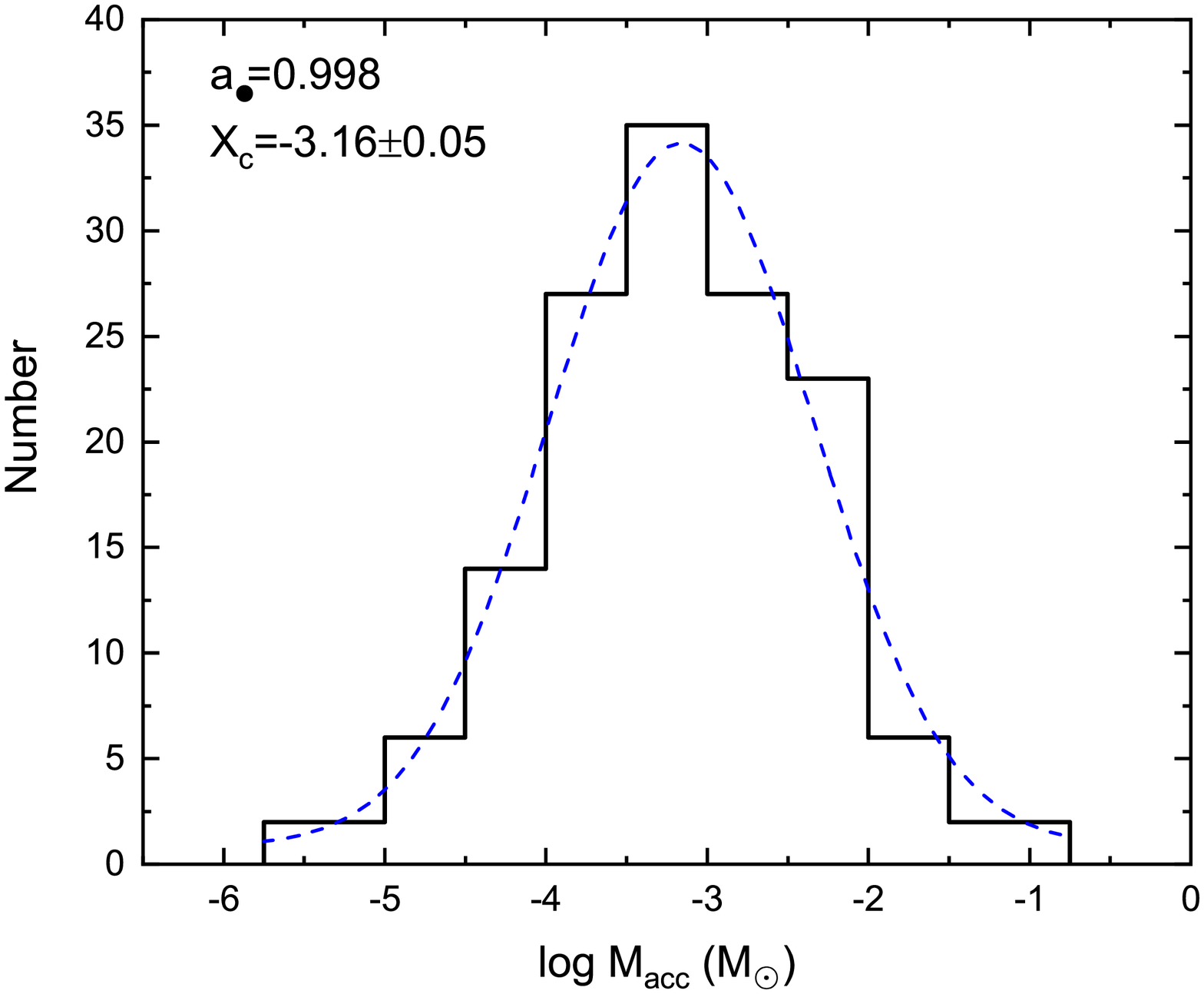}
\caption{Distributions of the accreted masses $M_{\rm acc}$ for different typical spins $a_\bullet$ with our selected GRB sample. The blue dashed line is the Guass fitting line.}
\end{figure*}

\clearpage
\begin{deluxetable}{ccccccccccccccccccccccccc}
\tabletypesize{\scriptsize}
\tablecaption{The parameters of GRBs with jet breaks. }
\tablewidth{0pt}
\tabletypesize{\scriptsize}

\tablehead{ \colhead{GRB}
&\colhead{$z$}
&\colhead{$T_{90}$}
&\colhead{$E_{\rm \gamma,iso}$}
&\colhead{$T_{\rm jet} $}
&\colhead{$\theta_{\rm jet}$}
&\colhead{Refs.\tablenotemark{a}}\\
& &[s] & [10$^{52}$ erg]&[day]&[rad]&&}

\startdata
970508	&	0.8349	&	20	&	0.63	$\pm$	0.13	&	25	$\pm$	5	&	0.371	$\pm$	0.028	&	1, 2, 2	\\
970828	&	0.9578	&	66	&	30.38	$\pm$	3.57	&	2.2	$\pm$	0.4	&	0.09	$\pm$	0.006	&	1, 2, 2	\\
980703	&	0.9662	&	400	&	7.42	$\pm$	0.71	&	3.4	$\pm$	0.5	&	0.126	$\pm$	0.007	&	3, 2, 2	\\
990123	&	1.6004	&	63	&	240.7	$\pm$	38.91	&	2.04	$\pm$	0.46	&	0.06	$\pm$	0.005	&	4, 2, 2	\\
990510	&	1.6187	&	68	&	18.1	$\pm$	2.72	&	1.2	$\pm$	0.08	&	0.068	$\pm$	0.002	&	5, 2, 2	\\
990705	&	0.8424	&	42	&	18.7	$\pm$	2.67	&	1	$\pm$	0.2	&	0.072	$\pm$	0.005	&	6, 2, 2	\\
990712	&	0.434	&	16.6	&	0.76	$\pm$	0.04	&	11.57			&	0.298			&	7, 2, 2	\\
991216	&	1.02	&	15	&	69.79	$\pm$	7.16	&	1.2	$\pm$	0.4	&	0.064	$\pm$	0.008	&	8, 2, 2	\\
000301C	&	2.03	&	10	&	199	$\pm$	35	&	6.52	$\pm$	0.22	&	0.09	$\pm$	0.001	&	9, 2, 2	\\
000926A	&	2.0379	&	55	&	27.1	$\pm$	5.9	&	1.8	$\pm$	0.1	&	0.072	$\pm$	0.001	&	1, 2, 2	\\
010222A	&	1.477	&	74	&	84.9	$\pm$	9.03	&	0.58	$\pm$	0.04	&	0.044	$\pm$	0.001	&	10, 2, 2	\\
010921	&	0.4509	&	20	&	0.97	$\pm$	0.1	&	33	$\pm$	6.5	&	0.426	$\pm$	0.031	&	1, 2, 2	\\
011121A	&	0.36	&	75	&	9.89	$\pm$	0.27	&	1.2	$\pm$	0.75	&	0.094	$\pm$	0.022	&	11,2,2	\\
011211	&	2.14	&	270	&	5.74	$\pm$	0.64	&	1.77	$\pm$	0.28	&	0.085	$\pm$	0.005	&	12, 2, 2	\\
020124	&	3.198	&	70	&	28.46	$\pm$	2.75	&	3	$\pm$	0.4	&	0.076	$\pm$	0.004	&	13, 2, 2	\\
020405	&	0.6899	&	60	&	10.64	$\pm$	0.89	&	1.67	$\pm$	0.52	&	0.097	$\pm$	0.011	&	14, 2, 2	\\
020813	&	1.254	&	125	&	68.35	$\pm$	1.71	&	0.43	$\pm$	0.06	&	0.042	$\pm$	0.002	&	15, 2, 2	\\
021004	&	2.332	&	78	&	3.47	$\pm$	0.46	&	7.6	$\pm$	0.3	&	0.153	$\pm$	0.002	&	1, 2, 2	\\
030226A	&	1.986	&	100	&	12.73	$\pm$	1.36	&	0.69	$\pm$	0.04	&	0.055	$\pm$	0.001	&	16, 2, 2	\\
030323	&	3.37	&	26	&	3.2	$\pm$	1	&	4.63			&	0.116			&	9, 2, 2	\\
030328A	&	1.5216	&	100	&	38.86	$\pm$	3.62	&	0.48	$\pm$	0.03	&	0.045	$\pm$	0.001	&	17, 2, 2	\\
030329	&	0.1685	&	50	&	1.55	$\pm$	0.15	&	0.47	$\pm$	0.05	&	0.089	$\pm$	0.003	&	9, 2, 2	\\
030429A	&	2.658	&	10	&	2.29	$\pm$	0.27	&	0.78	$\pm$	0.01	&	0.066	$\pm$	0	&	18, 2, 2	\\
050315	&	1.949	&	96	&	3.3	$\pm$	6.2	&	2.78	$\pm$	0.8	&	0.111	$\pm$	0.012	&	19, 2, 2	\\
050318A	&	1.44	&	32	&	2.3	$\pm$	0.16	&	0.24	$\pm$	0.12	&	0.05	$\pm$	0.009	&	20, 2, 2	\\
050319	&	3.2425	&	153	&	4.6	$\pm$	6.5	&	0.64	$\pm$	0.22	&	0.053	$\pm$	0.007	&	21, 2, 2	\\
050401	&	2.9	&	33	&	35	$\pm$	7	&	0.06	$\pm$	0.03	&	0.018	$\pm$	0.003	&	22, 2, 2	\\
050408	&	1.236	&	15	&	8.99			&	1.39	$\pm$	0.58	&	0.084	$\pm$	0.013	&	23, 2, 2	\\
050416A	&	0.654	&	2	&	0.1	$\pm$	0.01	&	0.01	$\pm$	0.01	&	0.026	$\pm$	0.01	&	24, 1, 25	\\
050502A	&	3.793	&	20	&	4	$\pm$	3	&	0.1	$\pm$	0.01	&	0.026	$\pm$	0.001	&	26, 2, 2	\\
050505A	&	4.2748	&	63	&	17.6	$\pm$	2.61	&	0.53	$\pm$	0.29	&	0.039	$\pm$	0.008	&	27, 2, 2	\\
050525A	&	0.606	&	9	&	2.3	$\pm$	0.49	&	0.16	$\pm$	0.09	&	0.05	$\pm$	0.011	&	28, 2, 2	\\
050730A	&	3.969	&	155	&	26	$\pm$	19	&	0.12	$\pm$	0.05	&	0.021	$\pm$	0.004	&	29, 2, 2	\\
050801	&	1.56	&	19	&	0.41	$\pm$	0.64	&	0.16	$\pm$	0.01	&	0.052	$\pm$	0.001	&	30, 2, 2	\\
050802	&	1.71	&	13	&	1.82	$\pm$	1.65	&	0.07	$\pm$	0.01	&	0.031	$\pm$	0.002	&	31, 2, 2	\\
050814	&	5.3	&	65	&	11.2	$\pm$	2.43	&	1.03	$\pm$	0.18	&	0.049	$\pm$	0.003	&	32, 2, 2	\\
050820A	&	2.62	&	50	&	103.36	$\pm$	8.23	&	7.52	$\pm$	2.31	&	0.097	$\pm$	0.011	&	33, 2, 2	\\
050826	&	0.3	&	35	&	0.01	$\pm$	0	&	0.45	$\pm$	0.39	&	0.159	$\pm$	0.052	&	34, 34, 25	\\
050904	&	6.29	&	225	&	133.36	$\pm$	13.89	&	2.6	$\pm$	1	&	0.048	$\pm$	0.007	&	35, 2, 2	\\
050922C	&	2.198	&	5	&	5.3	$\pm$	1.7	&	0.09		&	0.028		&	36, 2, 2	\\
051016B	&	0.94	&	4	&	0.1			&	1.56	$\pm$	0.1	&	0.162	$\pm$	0.004	&	37, 2, 2	\\
051022	&	0.8	&	154	&	56.04	$\pm$	5.33	&	2.9	$\pm$	0.2	&	0.095	$\pm$	0.002	&	1, 2, 2	\\
051109A	&	2.35	&	37	&	6.85	$\pm$	0.73	&	0.92	$\pm$	0.71	&	0.064	$\pm$	0.018	&	38, 2, 2	\\
051111	&	1.55	&	47	&	8.14			&	0.49	$\pm$	0.16	&	0.054	$\pm$	0.007	&	39, 25, 25	\\
060115	&	3.53	&	142	&	6.3	$\pm$	0.9	&	0.51	$\pm$	0.22	&	0.046	$\pm$	0.007	&	40, 2, 2	\\
060124	&	2.3	&	30	&	43.79	$\pm$	6.39	&	0.68	$\pm$	0.14	&	0.045	$\pm$	0.003	&	41, 2, 2	\\
060206A	&	4.048	&	7	&	4.3	$\pm$	0.9	&	0.6			&	0.049			&	42, 2, 2	\\
060210	&	3.91	&	220	&	32.23	$\pm$	1.84	&	0.3	$\pm$	0.1	&	0.03	$\pm$	0.004	&	43, 2, 2	\\
060218	&	0.03	&	2100	&	8.00E-04			&	0.82	$\pm$	1.53	&	0.294	$\pm$	0.206	&	44, 25, 25	\\
060418	&	1.49	&	52	&	13.55	$\pm$	2.71	&	0.07	$\pm$	0.04	&	0.025	$\pm$	0.005	&	45, 2, 2	\\
060526	&	3.22	&	14	&	2.75	$\pm$	0.37	&	2.41	$\pm$	0.06	&	0.094	$\pm$	0.001	&	46, 2, 2	\\
060605A	&	3.773	&	15	&	2.83	$\pm$	0.45	&	0.24	$\pm$	0.02	&	0.038	$\pm$	0.001	&	47, 2, 2	\\
060614	&	0.12	&	100	&	0.21	$\pm$	0.09	&	1.45	$\pm$	1.16	&	0.176	$\pm$	0.053	&	10, 2, 2	\\
060707	&	3.425	&	68	&	4.32	$\pm$	1.1	&	12.26	$\pm$	5.25	&	0.16	$\pm$	0.026	&	48, 2, 2	\\
060714	&	2.711	&	115	&	13.4	$\pm$	0.9	&	0.12	$\pm$	0.01	&	0.026	$\pm$	0.001	&	49, 2, 2	\\
060729	&	0.54	&	116	&	0.42	$\pm$	0.09	&	26.23	$\pm$	6.12	&	0.424	$\pm$	0.037	&	50, 2, 2	\\
060814	&	0.84	&	146	&	56.71	$\pm$	5.27	&	0.55	$\pm$	0.14	&	0.05	$\pm$	0.005	&	51, 2, 2	\\
060906	&	3.69	&	44	&	14.9	$\pm$	1.56	&	0.16	$\pm$	0.03	&	0.026	$\pm$	0.002	&	52, 2, 2	\\
060908	&	1.8836	&	18	&	7.18	$\pm$	1.91	&	0.01	$\pm$	0.02	&	0.012	$\pm$	0.009	&	10, 2, 2	\\
060926	&	3.21	&	7	&	1.15			&	0.06	$\pm$	0.05	&	0.026	$\pm$	0.008	&	53, 2, 2	\\
060927A	&	5.47	&	23	&	12.02	$\pm$	2.77	&	0.05			&	0.015			&	54, 2, 2	\\
061121	&	1.314	&	81	&	23.5	$\pm$	2.7	&	2.31			&	0.089			&	55, 2, 2	\\
061126	&	1.1588	&	191	&	31.42	$\pm$	3.59	&	1.52	$\pm$	0.23	&	0.075	$\pm$	0.004	&	56, 2, 2	\\
070125	&	1.5477	&	62	&	84.09	$\pm$	8.41	&	3.73	$\pm$	0.52	&	0.087	$\pm$	0.005	&	57, 2, 2	\\
070208	&	1.17	&	52	&	0.37			&	0.11	$\pm$	0.05	&	0.049	$\pm$	0.008	&	53, 2, 2	\\
070306	&	1.5	&	210	&	6	$\pm$	5	&	1.33	$\pm$	0.86	&	0.083	$\pm$	0.02	&	58, 2, 2	\\
070318	&	0.84	&	63	&	0.9	$\pm$	0.2	&	3.57	$\pm$	0.63	&	0.171	$\pm$	0.011	&	59, 2, 2	\\
070411	&	2.95	&	101	&	10	$\pm$	8	&	0.24	$\pm$	0.11	&	0.034	$\pm$	0.006	&	60, 2, 2	\\
070419A	&	0.97	&	116	&	0.24	$\pm$	0.1	&	0.02	$\pm$	0	&	0.027	$\pm$	0.002	&	61, 2, 2	\\
070508	&	0.82	&	23	&	8	$\pm$	2	&	0.58	$\pm$	0.93	&	0.066	$\pm$	0.04	&	10, 2, 2	\\
070611	&	2.04	&	12	&	0.92			&	1.13	$\pm$	0.39	&	0.092	$\pm$	0.012	&	62, 2, 2	\\
070714B	&	0.923	&	64	&	1.16	$\pm$	0.41	&	0.01	$\pm$	0	&	0.018	$\pm$	0.001	&	63, 10, 25	\\
070721B	&	3.626	&	340	&	36.5			&	0.11	$\pm$	0.01	&	0.021	$\pm$	0.001	&	64, 2, 2	\\
070810	&	2.17	&	11	&	1.73			&	0.09	$\pm$	0.04	&	0.032	$\pm$	0.005	&	65, 2, 2	\\
071003	&	1.1	&	148	&	20.6			&	0.41	$\pm$	0.07	&	0.049	$\pm$	0.003	&	66, 2, 2	\\
071010A	&	0.98	&	6	&	0.13			&	0.81	$\pm$	0.22	&	0.121	$\pm$	0.012	&	67, 2, 2	\\
071010B	&	0.947	&	36	&	2.32	$\pm$	0.4	&	3.44	$\pm$	0.39	&	0.146	$\pm$	0.006	&	68, 2, 2	\\
071031	&	2.69	&	180	&	3.9	$\pm$	0.6	&	0.71	$\pm$	0.35	&	0.06	$\pm$	0.011	&	69, 10, 25	\\
080210	&	2.64	&	45	&	5.13	$\pm$	2.13	&	0.14	$\pm$	0.06	&	0.031	$\pm$	0.005	&	70, 2, 2	\\
080310	&	2.43	&	365	&	20.9	$\pm$	2.1	&	0.34	$\pm$	0.04	&	0.038	$\pm$	0.002	&	71, 2, 2	\\
080319B	&	0.937	&	57	&	117.87			&	0.03	$\pm$	0.01	&	0.015	$\pm$	0.002	&	72, 2, 2	\\
080330A	&	1.5115	&	67	&	0.21	$\pm$	0.05	&	1			&	0.113			&	73, 2, 2	\\
080413A	&	2.433	&	46	&	7.83	$\pm$	3.55	&	0.01	$\pm$	0	&	0.012	$\pm$	0	&	74, 2, 2	\\
080413B	&	1.1	&	8	&	1.61	$\pm$	0.27	&	3.85	$\pm$	0.13	&	0.155	$\pm$	0.002	&	75, 2, 2	\\
080603A	&	1.688	&	150	&	2.2	$\pm$	0.8	&	1.16	$\pm$	0.46	&	0.087	$\pm$	0.013	&	74, 2, 2	\\
080710	&	0.85	&	120	&	1.68	$\pm$	0.22	&	0.23	$\pm$	0.02	&	0.057	$\pm$	0.001	&	76, 2, 2	\\
080810	&	3.35	&	108	&	45	$\pm$	5	&	0.11	$\pm$	0.02	&	0.02	$\pm$	0.002	&	77, 2, 2	\\
080928	&	1.692	&	280	&	2.82	$\pm$	1.17	&	0.14	$\pm$	0.04	&	0.038	$\pm$	0.004	&	78, 2, 2	\\
081007A    	&	0.5295	&	8	&	0.17			&	11.57			&	0.35			&	79, 2, 2	\\
081008A	&	1.967	&	185	&	6.92	$\pm$	1.67	&	0.21	$\pm$	0.07	&	0.038	$\pm$	0.005	&	80, 2, 2	\\
081203A	&	2.1	&	294	&	36.13	$\pm$	18.42	&	0.12	$\pm$	0.02	&	0.025	$\pm$	0.002	&	81, 2, 2	\\
090313A	&	3.375	&	78	&	3.2			&	1.04			&	0.066			&	82, 2, 2	\\
090323	&	3.568	&	133	&	410	$\pm$	50	&	17.8	$\pm$	19.6	&	0.103	$\pm$	0.043	&	10, 2, 2	\\
090328	&	0.7354	&	57	&	13	$\pm$	3	&	6.4	$\pm$	12	&	0.156	$\pm$	0.109	&	10, 2, 2	\\
090417B	&	0.345	&	2100	&	0.63			&	5.21	$\pm$	3.24	&	0.231	$\pm$	0.054	&	83, 2, 2	\\
090423	&	8.23	&	10	&	11	$\pm$	3	&	14.6	$\pm$	2.7	&	0.116	$\pm$	0.008	&	10, 2, 2	\\
090424A	&	0.544	&	49	&	4.6	$\pm$	0.9	&	11.57			&	0.231			&	10, 2, 2	\\
090618A	&	0.54	&	113	&	20			&	0.08	$\pm$	0.01	&	0.03	$\pm$	0.002	&	84, 2, 2	\\
090902B	&	1.8829	&	19	&	1.77	$\pm$	0.01	&	6.2	$\pm$	2.4	&	0.163	$\pm$	0.024	&	10, 2, 2	\\
090926A	&	2.1062	&	20	&	210	$\pm$	5.3	&	4.06	$\pm$	0.81	&	0.074	$\pm$	0.006	&	85, 2, 2	\\
091018	&	0.97	&	107	&	0.8	$\pm$	0.09	&	0.37	$\pm$	0.02	&	0.072	$\pm$	0.001	&	10, 2, 2	\\
091029A	&	2.752	&	39	&	7.4	$\pm$	0.74	&	0.03	$\pm$	0	&	0.017	$\pm$	0	&	86, 2, 2	\\
091127A	&	0.49	&	7	&	1.63	$\pm$	0.02	&	0.37	$\pm$	0.1	&	0.073	$\pm$	0.008	&	87, 2, 2	\\
091208B	&	1.063	&	15	&	2.01	$\pm$	0.07	&	3.59			&	0.148			&	88, 2, 2	\\
100219A	&	4.8	&	19	&	3.93			&	0.02	$\pm$	0	&	0.013	$\pm$	0.001	&	89, 2, 2	\\
100418	&	0.6239	&	7	&	0.1	$\pm$	0.06	&	4.2	$\pm$	1.1	&	0.251	$\pm$	0.025	&	90, 2, 2	\\
100814A	&	1.44	&	175	&	7.59	$\pm$	0.58	&	2	$\pm$	0.07	&	0.095	$\pm$	0.001	&	91, 2, 2	\\
100901A	&	1.408	&	439	&	6.3			&	11.57			&	0.188			&	92, 2, 2	\\
100906A	&	1.727	&	114	&	28.9	$\pm$	0.3	&	0.15	$\pm$	0.02	&	0.029	$\pm$	0.001	&	93, 2, 2	\\
110205A	&	2.22	&	257	&	48.38	$\pm$	6.38	&	1.2	$\pm$	0	&	0.056	$\pm$	0	&	94, 2, 2	\\
110801A	&	1.858	&	385	&	10.9	$\pm$	2.72	&	0.18	$\pm$	0.1	&	0.035	$\pm$	0.007	&	95, 2, 2	\\
111209A	&	0.677	&	25000	&	5.14	$\pm$	0.62	&	9.12	$\pm$	0.47	&	0.202	$\pm$	0.004	&	96, 2, 2	\\
120119A	&	1.728	&	254	&	27.2	$\pm$	3.63	&	0.13	$\pm$	0.02	&	0.028	$\pm$	0.002	&	97, 2, 2	\\
120326A	&	1.798	&	69	&	3.27	$\pm$	0.27	&	2.91	$\pm$	0.12	&	0.115	$\pm$	0.002	&	98, 2, 2	\\
120404A	&	2.876	&	39	&	9			&	0.06	$\pm$	0	&	0.021	$\pm$	0	&	99, 2, 2	\\
120729A	&	0.8	&	72	&	1.24	$\pm$	0.27	&	0.06	$\pm$	0.01	&	0.037	$\pm$	0.001	&	100, 2, 2	\\
121024A	&	2.298	&	69	&	2.51	$\pm$	1.56	&	0.54	$\pm$	0.17	&	0.059	$\pm$	0.007	&	101, 102, 102	\\
121027A	&	1.773	&	63	&	1.58	$\pm$	0.08	&	0.14			&	0.04			&	103, 2, 2	\\
121211A	&	1.023	&	182	&	0.63	$\pm$	0.54	&	0.44	$\pm$	0.28	&	0.078	$\pm$	0.019	&	104, 2, 2	\\
130427A	&	0.3399	&	163	&	77.01	$\pm$	7.88	&	0.43	$\pm$	0.05	&	0.05	$\pm$	0.002	&	105, 2, 2	\\
130427B	&	2.78	&	27	&	3.16	$\pm$	1.75	&	0.21	$\pm$	0.09	&	0.038	$\pm$	0.006	&	106, 2, 2	\\
130606A	&	5.91	&	277	&	28.3	$\pm$	5.2	&	0.17	$\pm$	0.05	&	0.022	$\pm$	0.002	&	107, 2, 2	\\
130907A	&	1.238	&	214	&	300			&	0.25	$\pm$	0.02	&	0.028	$\pm$	0.001	&	9, 2, 2	\\
131030A	&	1.293	&	41	&	32.7	$\pm$	1.3	&	2.91	$\pm$	0.56	&	0.093	$\pm$	0.007	&	108, 2, 2	\\
140311A	&	4.954	&	71	&	10			&	1.3	$\pm$	1.1	&	0.056	$\pm$	0.018	&	109, 2, 2	\\
140512A	&	0.725	&	155	&	7.25	$\pm$	0.61	&	0.21	$\pm$	0.02	&	0.047	$\pm$	0.001	&	110, 2, 2	\\
140629A	&	2.276	&	42	&	4.4			&	0.42	$\pm$	0.09	&	0.05	$\pm$	0.004	&	111, 2, 2	\\
141121A	&	1.469	&	550	&	8			&	3.8	$\pm$	0.5	&	0.119	$\pm$	0.006	&	112, 2, 2	\\
151027A	&	0.81	&	130	&	3.3	$\pm$	0.41	&	0.69	$\pm$	0.21	&	0.079	$\pm$	0.009	&	113, 2, 2	\\
160131A	&	0.97	&	325	&	87	$\pm$	6.6	&	0.13	$\pm$	0.02	&	0.027	$\pm$	0.001	&	114, 2, 2	\\
160227A	&	2.38	&	317	&	5.56	$\pm$	0.36	&	1	$\pm$	0.06	&	0.067	$\pm$	0.002	&	115, 2, 2	\\
160410A	&	1.717	&	8	&	4			&	0.02	$\pm$	0.02	&	0.017	$\pm$	0.008	&	116, 117, 117	\\
160509A	&	1.17	&	371	&	86			&	3.7	$\pm$	0.8	&	0.092	$\pm$	0.007	&	9, 2, 2	\\
160625B	&	1.406	&	460	&	300			&	22	$\pm$	4	&	0.148	$\pm$	0.01	&	9, 2, 2	\\
161017A	&	2.013	&	216	&	6.87	$\pm$	0.72	&	0.57	$\pm$	1.61	&	0.055	$\pm$	0.059	&	118, 2, 2	\\
161219B	&	0.1475	&	7	&	0.01			&	17.4	$\pm$	8.78	&	0.636	$\pm$	0.12	&	119, 117, 117	\\
170405A	&	3.51	&	165	&	9.01			&	0.06	$\pm$	0.01	&	0.02	$\pm$	0.001	&	120, 2, 2	\\
170519A	&	0.818	&	216	&	0.19			&	0.54			&	0.103			&	121, 117, 117	\\
171010A	&	0.33	&	117	&	22			&	6.48	$\pm$	27.09	&	0.162	$\pm$	0.254	&	7, 2, 2	\\
180115A	&	2.487	&	41	&	1.04			&	0.2	$\pm$	0.04	&	0.045	$\pm$	0.003	&	122, 2, 2	\\
180620B	&	1.1175	&	199	&	3.04	$\pm$	0.03	&	2.8	$\pm$	0.44	&	0.127	$\pm$	0.007	&	1, 2, 2	\\
180720B	&	0.654	&	69	&	33.97	$\pm$	0.01	&	1.08	$\pm$	0.27	&	0.072	$\pm$	0.007	&	1, 2, 2	\\
180728A	&	0.117	&	9	&	0.3			&	1.82	$\pm$	1.08	&	0.183	$\pm$	0.041	&	1, 2, 2	\\
181010A	&	1.39	&	16	&	3.4			&	0.89	$\pm$	0.96	&	0.078	$\pm$	0.032	&	123, 117, 117	\\
181020A	&	2.938	&	238	&	82.8	$\pm$	11.6	&	0.13	$\pm$	0.04	&	0.021	$\pm$	0.003	&	124, 2, 2	\\
190114C	&	0.42	&	362	&	27.03	$\pm$	0.24	&	0.1	$\pm$	0.04	&	0.032	$\pm$	0.005	&	125, 2, 2	\\
	
\enddata
\tablenotetext{a}{References for $T_{90}$, $E_{\gamma,iso}$ and $T_{jet}$, respectively.}
\tablerefs{(1) Minaev et al. 2020; (2) Zhao et al. 2020; (3) Kong et al. 2009; (4) Maiorano et al. 2005; (5) Wijers et al. 1999; (6) Wang et al. 2003; (7) Xue et al. 2019; (8) Sagar et al. 2000; (9) Ruffini et al. 2018 (10) Song et al. 2018; (11) Piro et al. 2005; (12) Holland et al. 2002; (13) Berger et al. 2002; (14) Stratta et al. 2005; (15) Sato et al. 2005; (16) Dai et al. 2003; (17) Gorosabel et al. 2005; (18) Jakobsson et al. 2004; (19) Krimm et al. 2005a; (20) Krimm et al. 2005b; (21) Krimm et al. 2005c; (22) Sakamoto et al. 2005a; (23) Capalbi et al. 2007; (24) Soderberg et al. 2007; (25) Lu et al. 2012; (26) Guidorzi et al. 2005; (27) Hurkett et al. 2006; (28) Cummings et al. 2005a; (29) Perri et al. 2007; (30) Sakamoto et al. 2005b; (31) Palmer et al. 2005; (32) Tueller et al. 2005; (33) Cummings et al. 2005b; (34) Mirabal et al. 2007; (35) Gou et al. 2007; (36) Krimm et al. 2005d; (37) Barbier et al. 2005; (38) Fenimore et al. 2005; (39) Krimm et al. 2005e; (40) Barbier et al. 2006; (41) Holland et al. 2006; (42) Liu et al. 2008; (43) Curran et al. 2007;(44) Samuelsson et al. 2020; (45) Cummings et al. 2006; (46) Markwardt et al. 2006; (47) Ferrero et al. 2009; (48) Stamatikos et al. 2006a; (49) Krimm et al. 2007a; (50) Parsons et al. 2006; (51) Stamatikos et al. 2006b; (52) Sakamoto et al. 2006; (53) Butler et al. 2007; (54) Stamatikos et al. 2006c; (55) Fenimore et al. 2006; (56) Krimm et al. 2006; (57) Bellm et al. 2008; (58) Jaunsen et al. 2008; (59) Cummings et al. 2007; (60) Markwardt et al. 2007a; (61) Stamatikos et al. 2007a; (62) Barbier et al. 2007a; (63) Barbier et al. 2007b; (64) Palmer et al. 2007; (65) Markwardt et al. 2007b; (66) Ukwatta et al.2007; (67) Krimm et al. 2007b; (68) Markwardt et al. 2007c; (69) Stamatikos et al. 2007b; (70) Ukwatta et al. 2008a; (71) Tueller et al. 2008a; (72) Cummings et al. 2008; (73) Markwardt et al. 2008; (74) Tueller et al. 2008b; (75) Barthelmy et al. 2008; (76) Tueller et al. 2008c; (77) Sakamoto et al. 2008; (78) Fenimore et al. 2008; (79) Jin et al. 2013; (80) Palmer et al. 2008; (81) Ukwatta et al. 2008b; (82) Sakamoto et al. 2009; (83) Holland et al.2010; (84) Baumgartner et al. 2009a; (85) Swenson et al. 2010; (86) Barthelmy et al. 2009; (87) Stamatikos et al. 2009; (88) Baumgartner et al. 2009b; (89) Baumgartner et al. 2010; (90) de Ugarte et al. 2018; (91) Krimm et al. 2010; (92) Hartoog et al. 2013; (93) Gorbovskoy et al. 2012; (94) Zheng et al. 2012; (95) Barthelmy et al. 2011; (96) Gendre et al. 2013; (97) Stamatikos et al. 2012; (98) Melandri et al. 2014; (99) Ukwatta et al. 2012; (100) Palmer et al. 2012; (101) Barthelmy et al. 2012a; (102) Xi et al. 2017; (103) Barthelmy et al. 2012b; (104) Barthelmy et al. 2012c; (105) Becerra et al. 2017; (106) Barthelmy et al. 2013a; (107) Barthelmy et al. 2013b; (108) Huang et al. 2017; (109) Krimm et al. 2014a; (110) Sakamoto et al. 2014; (111) Hu et al. 2019; (112) Krimm et al. 2014b; (113) Palmer et al. 2015; (114) Cummings et al. 2016a; (115) Sakamoto et al. 2016a; (116) Sakamoto et al. 2016b; (117) this work; (118) Cummings et al. 2016b; (119) Palmer et al. 2016; (120) Palmer et al. 2017; (121) Krimm et al. 2017; (122) Barthelmy et al. 2018; (123) Lien et al. 2018; (124) Sakamoto et al. 2018; (125) Krimm et al. 2019. }
\end{deluxetable}


\clearpage
\begin{deluxetable}{ccccccccccccccccccccccccc}
\tabletypesize{\scriptsize}
\tablecaption{The parameters of GRBs with jet breaks. }
\tablewidth{0pt}
\tabletypesize{\scriptsize}

\tablehead{ \colhead{GRB}
&\colhead{$L_{\gamma,iso}$}
&\colhead{$L_{\gamma}$}
&\colhead{$\dot{m}_{\rm in}(a_\bullet=0.5) $}
&\colhead{$\dot{m}_{\rm in}(a_\bullet=0.7) $}
&\colhead{$\dot{m}_{\rm in}(a_\bullet=0.87) $}
&\colhead{$\dot{m}_{\rm in}(a_\bullet=0.998) $}\\
&[10$^{51}$ erg/s] & [10$^{49}$ erg/s]&$[10^{-4}M_{\odot}~\rm s^{-1}]$&$[10^{-4}M_{\odot}~\rm s^{-1}]$&$[10^{-4}M_{\odot}~\rm s^{-1}]$&$[10^{-4}M_{\odot}~\rm s^{-1}]$&}

\startdata
    970508 & 0.580  & 3.984  & 42.300  & 17.027  & 7.504  & 2.228  \\
    970828 & 8.987  & 3.609  & 38.367  & 15.444  & 6.806  & 2.021  \\
    980703 & 0.365  & 0.288  & 3.061  & 1.232  & 0.543  & 0.161  \\
    990123 & 98.882  & 18.074  & 192.368  & 77.433  & 34.126  & 10.132  \\
    990510 & 6.972  & 1.626  & 17.324  & 6.973  & 3.073  & 0.912  \\
    990705 & 8.204  & 2.155  & 22.850  & 9.198  & 4.054  & 1.204  \\
    990712 & 0.655  & 2.903  & 30.821  & 12.406  & 5.468  & 1.623  \\
    991216 & 92.949  & 18.795  & 199.808  & 80.427  & 35.446  & 10.524  \\
    000301C & 602.970  & 246.211  & 2614.510  & 1052.399  & 463.810  & 137.710  \\
    000926A & 15.061  & 3.850  & 40.918  & 16.470  & 7.259  & 2.155  \\
    010222A & 28.418  & 2.722  & 28.908  & 11.636  & 5.128  & 1.523  \\
    010921 & 0.706  & 6.406  & 68.126  & 27.422  & 12.085  & 3.588  \\
    011121A & 1.793  & 0.795  & 8.449  & 3.401  & 1.499  & 0.445  \\
    011211 & 0.667  & 0.242  & 2.572  & 1.035  & 0.456  & 0.135  \\
   020124 & 17.066  & 4.960  & 52.715  & 21.219  & 9.352  & 2.777  \\
    020405 & 2.996  & 1.420  & 15.092  & 6.075  & 2.677  & 0.795  \\
    020813 & 12.326  & 1.069  & 11.372  & 4.578  & 2.017  & 0.599  \\
    021004 & 1.482  & 1.742  & 18.493  & 7.444  & 3.281  & 0.974  \\
    030226A & 3.802  & 0.579  & 6.154  & 2.477  & 1.092  & 0.324  \\
    030323 & 5.378  & 3.627  & 38.580  & 15.529  & 6.844  & 2.032  \\
    030328A & 9.799  & 0.977  & 10.384  & 4.180  & 1.842  & 0.547  \\
    030329 & 0.362  & 0.143  & 1.520  & 0.612  & 0.270  & 0.080  \\
    030429A & 8.123  & 1.790  & 19.024  & 7.658  & 3.375  & 1.002  \\
    050315 & 1.018  & 0.624  & 6.632  & 2.670  & 1.176  & 0.349  \\
    050318A & 1.755  & 0.216  & 2.296  & 0.924  & 0.407  & 0.121  \\
    050319 & 1.280  & 0.183  & 1.945  & 0.783  & 0.345  & 0.102  \\
    050401 & 40.991  & 0.636  & 6.759  & 2.721  & 1.199  & 0.356  \\
    050408 & 13.401  & 4.680  & 49.739  & 20.021  & 8.824  & 2.620  \\
    050416A & 0.689  & 0.023  & 0.244  & 0.098  & 0.043  & 0.013  \\
    050502A & 9.586  & 0.315  & 3.348  & 1.348  & 0.594  & 0.176  \\
    050505A & 14.736  & 1.109  & 11.797  & 4.749  & 2.093  & 0.621  \\
    050525A & 4.195  & 0.522  & 5.548  & 2.233  & 0.984  & 0.292  \\
    050730A & 8.335  & 0.192  & 2.041  & 0.821  & 0.362  & 0.107  \\
    050801 & 0.541  & 0.074  & 0.792  & 0.319  & 0.140  & 0.042  \\
    050802 & 3.793  & 0.182  & 1.934  & 0.779  & 0.343  & 0.102  \\
    050814 & 10.855  & 1.318  & 14.029  & 5.647  & 2.489  & 0.739  \\
    050820A & 74.830  & 35.075  & 373.046  & 150.159  & 66.178  & 19.649  \\
    050826 & 0.003  & 0.004  & 0.045  & 0.018  & 0.008  & 0.002  \\
    050904 & 43.210  & 5.068  & 53.884  & 21.690  & 9.559  & 2.838  \\
    050922C & 33.899  & 1.326  & 14.135  & 5.690  & 2.508  & 0.745  \\
    051016B & 0.470  & 0.618  & 6.568  & 2.644  & 1.165  & 0.346  \\
    051022 & 6.562  & 2.962  & 31.459  & 12.663  & 5.581  & 1.657  \\
    051109A & 6.164  & 1.249  & 13.285  & 5.348  & 2.357  & 0.700  \\
    051111 & 4.416  & 0.655  & 6.961  & 2.802  & 1.235  & 0.367  \\
    060115 & 2.010  & 0.213  & 2.264  & 0.911  & 0.402  & 0.119  \\
    060124 & 48.169  & 4.946  & 52.609  & 21.176  & 9.333  & 2.771  \\
    060206A & 31.009  & 3.765  & 40.068  & 16.128  & 7.108  & 2.110  \\
    060210 & 7.193  & 0.320  & 3.401  & 1.369  & 0.603  & 0.179  \\
    060218 & 0.000  & 0.000  & 0.000  & 0.000  & 0.000  & 0.000  \\
    060418 & 6.488  & 0.201  & 2.136  & 0.860  & 0.379  & 0.113  \\
    060526 & 8.422  & 3.710  & 39.430  & 15.872  & 6.995  & 2.077  \\
    060605A & 9.005  & 0.637  & 6.770  & 2.725  & 1.201  & 0.357  \\
    060614 & 0.024  & 0.036  & 0.387  & 0.156  & 0.069  & 0.020  \\
    060707 & 2.811  & 3.616  & 38.474  & 15.487  & 6.825  & 2.026  \\
    060714 & 4.324  & 0.149  & 1.584  & 0.637  & 0.281  & 0.083  \\
    060729 & 0.056  & 0.505  & 5.367  & 2.160  & 0.952  & 0.283  \\
    060814 & 7.147  & 0.909  & 9.661  & 3.889  & 1.714  & 0.509  \\
    060906 & 16.028  & 0.559  & 5.941  & 2.391  & 1.054  & 0.313  \\
    060908 & 11.496  & 0.087  & 0.921  & 0.371  & 0.163  & 0.049  \\
    060926 & 6.819  & 0.235  & 2.498  & 1.005  & 0.443  & 0.132  \\
    060927A & 34.416  & 0.393  & 4.177  & 1.681  & 0.741  & 0.220  \\
    061121 & 6.714  & 2.630  & 27.952  & 11.251  & 4.959  & 1.472  \\
    061126 & 3.551  & 0.994  & 10.564  & 4.252  & 1.874  & 0.556  \\
    070125 & 34.442  & 13.075  & 139.228  & 56.042  & 24.699  & 7.333  \\
    070208 & 0.153  & 0.018  & 0.192  & 0.077  & 0.034  & 0.010  \\
    070306 & 0.714  & 0.246  & 2.615  & 1.052  & 0.464  & 0.138  \\
    070318 & 0.263  & 0.383  & 4.071  & 1.638  & 0.722  & 0.214  \\
    070411 & 3.911  & 0.232  & 2.466  & 0.993  & 0.437  & 0.130  \\
    070419A & 0.041  & 0.001  & 0.015  & 0.006  & 0.003  & 0.001  \\
    070508 & 6.222  & 1.355  & 14.454  & 5.818  & 2.564  & 0.761  \\
    070611 & 2.305  & 0.966  & 10.267  & 4.133  & 1.821  & 0.541  \\
    070714B & 0.349  & 0.006  & 0.060  & 0.024  & 0.011  & 0.003  \\
    070721B & 4.966  & 0.106  & 1.127  & 0.453  & 0.200  & 0.059  \\
    070810 & 4.986  & 0.260  & 2.763  & 1.112  & 0.490  & 0.146  \\
    071003 & 2.884  & 0.343  & 3.645  & 1.467  & 0.647  & 0.192  \\
    071010A & 0.429  & 0.316  & 3.358  & 1.352  & 0.596  & 0.177  \\
    071010B & 1.265  & 1.356  & 14.454  & 5.818  & 2.564  & 0.761  \\
    071031 & 0.800  & 0.143  & 1.520  & 0.612  & 0.270  & 0.080  \\
    080210 & 4.150  & 0.202  & 2.147  & 0.864  & 0.381  & 0.113  \\
    080310 & 1.964  & 0.140  & 1.488  & 0.599  & 0.264  & 0.078  \\
    080319B & 40.054  & 0.461  & 4.900  & 1.972  & 0.869  & 0.258  \\
    080330A & 0.086  & 0.055  & 0.588  & 0.237  & 0.104  & 0.031  \\
    080413A & 5.844  & 0.042  & 0.449  & 0.181  & 0.080  & 0.024  \\
    080413B & 4.236  & 5.112  & 54.310  & 21.861  & 9.634  & 2.861  \\
    080603A & 0.394  & 0.149  & 1.584  & 0.637  & 0.281  & 0.083  \\
    080710 & 0.259  & 0.042  & 0.444  & 0.179  & 0.079  & 0.023  \\
    080810 & 18.125  & 0.375  & 3.986  & 1.604  & 0.707  & 0.210  \\
    080928 & 0.271  & 0.020  & 0.210  & 0.085  & 0.037  & 0.011  \\
    \multicolumn{1}{l}{081007A    } & 0.329  & 2.012  & 21.362  & 8.599  & 3.790  & 1.125  \\
    081008A & 1.110  & 0.081  & 0.859  & 0.346  & 0.152  & 0.045  \\
    081203A & 3.810  & 0.115  & 1.222  & 0.492  & 0.217  & 0.064  \\
    090313A & 1.772  & 0.390  & 4.145  & 1.668  & 0.735  & 0.218  \\
    090323 & 140.712  & 74.908  & 796.044  & 320.425  & 141.217  & 41.929  \\
   090328 & 3.958  & 4.791  & 50.909  & 20.492  & 9.031  & 2.681  \\
    090417B & 0.004  & 0.011  & 0.113  & 0.045  & 0.020  & 0.006  \\
    090423 & 98.573  & 65.939  & 700.391  & 281.923  & 124.248  & 36.891  \\
    090424A & 1.436  & 3.836  & 40.812  & 16.428  & 7.240  & 2.150  \\
    090618A & 2.721  & 0.122  & 1.297  & 0.522  & 0.230  & 0.068  \\
    090902B & 2.640  & 3.511  & 37.305  & 15.016  & 6.618  & 1.965  \\
    090926A & 326.151  & 90.502  & 961.842  & 387.163  & 170.629  & 50.662  \\
    091018 & 0.148  & 0.039  & 0.413  & 0.166  & 0.073  & 0.022  \\
    091029A & 7.083  & 0.105  & 1.116  & 0.449  & 0.198  & 0.059  \\
    091127A & 3.421  & 0.921  & 9.788  & 3.940  & 1.736  & 0.516  \\
    091208B & 2.783  & 3.057  & 32.522  & 13.091  & 5.769  & 1.713  \\
    100219A & 12.124  & 0.109  & 1.158  & 0.466  & 0.206  & 0.061  \\
    100418 & 0.230  & 0.721  & 7.663  & 3.084  & 1.359  & 0.404  \\
    100814A & 1.061  & 0.476  & 5.059  & 2.036  & 0.897  & 0.266  \\
    100901A & 0.346  & 0.611  & 6.494  & 2.614  & 1.152  & 0.342  \\
    100906A & 6.889  & 0.296  & 3.146  & 1.266  & 0.558  & 0.166  \\
    110205A & 6.061  & 0.947  & 10.065  & 4.051  & 1.785  & 0.530  \\
    110801A & 0.809  & 0.049  & 0.522  & 0.210  & 0.093  & 0.027  \\
    111209A & 0.003  & 0.007  & 0.075  & 0.030  & 0.013  & 0.004  \\
    120119A & 2.923  & 0.113  & 1.201  & 0.483  & 0.213  & 0.063  \\
    120326A & 1.325  & 0.875  & 9.300  & 3.743  & 1.650  & 0.490  \\
    120404A & 9.014  & 0.197  & 2.094  & 0.843  & 0.371  & 0.110  \\
    120729A & 0.312  & 0.021  & 0.224  & 0.090  & 0.040  & 0.012  \\
    121024A & 1.200  & 0.212  & 2.253  & 0.907  & 0.400  & 0.119  \\
    121027A & 0.700  & 0.057  & 0.607  & 0.244  & 0.108  & 0.032  \\
    121211A & 0.070  & 0.021  & 0.229  & 0.092  & 0.041  & 0.012  \\
    130427A & 6.331  & 0.791  & 8.407  & 3.384  & 1.491  & 0.443  \\
    130427B & 4.424  & 0.327  & 3.475  & 1.399  & 0.617  & 0.183  \\
    130606A & 7.070  & 0.167  & 1.775  & 0.714  & 0.315  & 0.093  \\
    130907A & 31.374  & 1.258  & 13.391  & 5.390  & 2.376  & 0.705  \\
    131030A & 18.244  & 7.869  & 83.643  & 33.668  & 14.838  & 4.406  \\
    140311A & 8.339  & 1.294  & 13.710  & 5.519  & 2.432  & 0.722  \\
    140512A & 0.808  & 0.089  & 0.942  & 0.379  & 0.167  & 0.050  \\
    140629A & 3.432  & 0.437  & 4.644  & 1.870  & 0.824  & 0.245  \\
    141121A & 0.359  & 0.255  & 2.710  & 1.091  & 0.481  & 0.143  \\
    151027A & 0.461  & 0.144  & 1.530  & 0.616  & 0.272  & 0.081  \\
    160131A & 5.274  & 0.196  & 2.083  & 0.839  & 0.370  & 0.110  \\
    160227A & 0.594  & 0.134  & 1.424  & 0.573  & 0.253  & 0.075  \\
    160410A & 13.254  & 0.192  & 2.041  & 0.821  & 0.362  & 0.107  \\
    160509A & 5.030  & 2.129  & 22.638  & 9.112  & 4.016  & 1.192  \\
    160625B & 15.691  & 17.124  & 181.740  & 73.155  & 32.240  & 9.573  \\
    161017A & 0.957  & 0.146  & 1.552  & 0.625  & 0.275  & 0.082  \\
    161219B & 0.019  & 0.388  & 4.124  & 1.660  & 0.732  & 0.217  \\
    170405A & 2.467  & 0.050  & 0.530  & 0.213  & 0.094  & 0.028  \\
    170519A & 0.016  & 0.008  & 0.088  & 0.036  & 0.016  & 0.005  \\
    171010A & 2.508  & 3.279  & 34.860  & 14.032  & 6.184  & 1.836  \\
    180115A & 0.887  & 0.088  & 0.934  & 0.376  & 0.166  & 0.049  \\
    180620B & 0.324  & 0.261  & 2.774  & 1.117  & 0.492  & 0.146  \\
    180720B & 8.197  & 2.131  & 22.638  & 9.112  & 4.016  & 1.192  \\
    180728A & 0.386  & 0.649  & 6.898  & 2.776  & 1.224  & 0.363  \\
    181010A & 4.955  & 1.509  & 16.048  & 6.460  & 2.847  & 0.845  \\
    181020A & 13.700  & 0.302  & 3.210  & 1.292  & 0.569  & 0.169  \\
    190114C & 1.062  & 0.054  & 0.575  & 0.231  & 0.102  & 0.030  \\

\enddata

\end{deluxetable}


\begin{thebibliography}{}
\bibitem[Abdo et al.(2011)]{2011ApJ...734L..27A} Abdo, A.~A., Ackermann, M., Ajello, M., et al.\ 2011, \apjl, 734, L27
\bibitem[Barbier et al.(2007)]{2007GCN..6502....1B} Barbier, L., Barthelmy, S.~D., Cummings, J., et al.\ 2007a, GRB Coordinates Network, Circular Service, No. 6502, \#1 (2007), 6502
\bibitem[Barbier et al.(2007)]{2007GCN..6623....1B} Barbier, L., Barthelmy, S.~D., Cummings, J., et al.\ 2007b, GRB Coordinates Network, Circular Service, No. 6623, \#1 (2007), 6623
\bibitem[Barbier et al.(2006)]{2006GCN..4518....1B} Barbier, L., Barthelmy, S., Cummings, J., et al.\ 2006, GRB Coordinates Network, Circular Service, No. 4518, \#1 (2006), 4518
\bibitem[Barbier et al.(2005)]{2005GCN..4104....1B} Barbier, L., Barthelmy, S., Cummings, J., et al.\ 2005, GRB Coordinates Network, Circular Service, No. 4104, \#1 (2005), 4104
\bibitem[Barthelmy et al.(2012)]{2012GCN.13899....1B} Barthelmy, S.~D., Sakamoto, T., Baumgartner, W.~H., et al.\ 2012a, GRB Coordinates Network, Circular Service, No. 13899, \#1 (2012), 13899
\bibitem[Barthelmy et al.(2012)]{2012GCN.13910....1B} Barthelmy, S.~D., Baumgartner, W.~H., Cummings, J.~R., et al.\ 2012b, GRB Coordinates Network, Circular Service, No. 13910, \#1 (2012), 13910
\bibitem[Barthelmy et al.(2012)]{2012GCN.14067....1B} Barthelmy, S.~D., Baumgartner, W.~H., Cummings, J.~R., et al.\ 2012c, GRB Coordinates Network, Circular Service, No. 14067, \#1 (2012), 14067
\bibitem[Barthelmy et al.(2013)]{2013GCN.14469....1B} Barthelmy, S.~D., Baumgartner, W.~H., Cummings, J.~R., et al.\ 2013a, GRB Coordinates Network, Circular Service, No. 14469, \#1 (2013), 14469
\bibitem[Barthelmy et al.(2013)]{2013GCN.14819....1B} Barthelmy, S.~D., Baumgartner, W.~H., Cummings, J.~R., et al.\ 2013b, GRB Coordinates Network, Circular Service, No. 14819, \#1 (2013), 14819
\bibitem[Barthelmy et al.(2011)]{2011GCN.12237....1B} Barthelmy, S.~D., Baumgartner, W.~H., Cummings, J.~R., et al.\ 2011, GRB Coordinates Network, Circular Service, No. 12237, \#1 (2011), 12237
\bibitem[Barthelmy et al.(2009)]{2009GCN.10103....1B} Barthelmy, S.~D., Baumgartner, W.~H., Cummings, J.~R., et al.\ 2009, GRB Coordinates Network, Circular Service, No. 10103, \#1 (2009), 10103
\bibitem[Barthelmy et al.(2018)]{2018GCN.22348....1B} Barthelmy, S.~D., Cannizzo, J.~K., Cummings, J.~R., et al.\ 2018, GRB Coordinates Network, Circular Service, No. 22348, \#1 (2018/January-0), 22348
\bibitem[Barthelmy et al.(2008)]{2008GCN..7606....1B} Barthelmy, S.~D., Baumgartner, W., Cummings, J., et al.\ 2008, GRB Coordinates Network, Circular Service, No. 7606, \#1 (2008), 7606
\bibitem[Baumgartner et al.(2010)]{2010GCN.10434....1B} Baumgartner, W.~H., Barthelmy, S.~D., Cummings, J.~R., et al.\ 2010, GRB Coordinates Network, Circular Service, No. 10434, \#1 (2010), 10434
\bibitem[Baumgartner et al.(2009)]{2009GCN..9530....1B} Baumgartner, W.~H., Barthelmy, S.~D., Cummings, J.~R., et al.\ 2009a, GRB Coordinates Network, Circular Service, No. 9530, \#1 (2009), 9530
\bibitem[Baumgartner et al.(2009)]{2009GCN.10265....1B} Baumgartner, W.~H., Barthelmy, S.~D., Cummings, J.~R., et al.\ 2009b, GRB Coordinates Network, Circular Service, No. 10265, \#1 (2009), 10265
\bibitem[Becerra et al.(2017)]{2017ApJ...837..116B} Becerra, R.~L., Watson, A.~M., Lee, W.~H., et al.\ 2017, \apj, 837, 116
\bibitem[Bellm et al.(2008)]{2008ApJ...688..491B} Bellm, E.~C., Hurley, K., Pal'shin, V., et al.\ 2008, \apj, 688, 491
\bibitem[Berger et al.(2002)]{2002ApJ...581..981B} Berger, E., Kulkarni, S.~R., Bloom, J.~S., et al.\ 2002, \apj, 581, 981
\bibitem[Blandford \& Znajek(1977)]{1977MNRAS.179..433B} Blandford, R.~D. \& Znajek, R.~L.\ 1977, \mnras, 179, 433
\bibitem[Bucciantini et al.(2009)]{2009MNRAS.396.2038B} Bucciantini, N., Quataert, E., Metzger, B.~D., et al.\ 2009, \mnras, 396, 2038
\bibitem[Burrows et al.(2005)]{2005Sci...309.1833B} Burrows, D.~N., Romano, P., Falcone, A., et al.\ 2005, Science, 309, 1833
\bibitem[Butler et al.(2007)]{2007ApJ...671..656B} Butler, N.~R., Kocevski, D., Bloom, J.~S., et al.\ 2007, \apj, 671, 656
\bibitem[Capalbi et al.(2007)]{2007A&A...462..913C} Capalbi, M., Malesani, D., Perri, M., et al.\ 2007, \aap, 462, 913
\bibitem[Chen \& Beloborodov (2007)]{2007ApJ...657..383C}Chen, W.-X., \& Beloborodov, A. M. \ 2007, \apj, 657, 383
\bibitem[Chincarini et al.(2007)]{2007ApJ...671.1903C} Chincarini, G., Moretti, A., Romano, P., et al.\ 2007, \apj, 671, 1903
\bibitem[Chincarini et al.(2010)]{2010MNRAS.406.2113C} Chincarini, G., Mao, J., Margutti, R., et al.\ 2010, \mnras, 406, 2113
\bibitem[Cummings et al.(2005)]{2005GCN..3479....1C} Cummings, J., Barbier, L., Barthelmy, S., et al.\ 2005a, GRB Coordinates Network, Circular Service, No. 3479, \#1 (2005), 3479
\bibitem[Cummings et al.(2005)]{2005GCN..3858....1C} Cummings, J., Barbier, L., Barthelmy, S., et al.\ 2005b, GRB Coordinates Network, Circular Service, No. 3858, \#1 (2005), 3858
\bibitem[Cummings et al.(2006)]{2006GCN..4975....1C} Cummings, J., Barbier, L., Barthelmy, S., et al.\ 2006, GRB Coordinates Network, Circular Service, No. 4975, \#1 (2006), 4975
\bibitem[Cummings et al.(2007)]{2007GCN..6733....1C} Cummings, J.~R., Barthelmy, S.~D., Chester, M.~M., et al.\ 2007, GRB Coordinates Network, Circular Service, No. 6733, \#1 (2007), 6733
\bibitem[Cummings et al.(2016)]{2016GCN.18959....1C} Cummings, J.~R., Barthelmy, S.~D., Gehrels, N., et al.\ 2016a, GRB Coordinates Network, Circular Service, No. 18959, \#1 (2016), 18959
\bibitem[Cummings et al.(2016)]{2016GCN.20076....1C} Cummings, J.~R., Barthelmy, S.~D., Gehrels, N., et al.\ 2016b, GRB Coordinates Network, Circular Service, No. 20076, \#1 (2016), 20076
\bibitem[Cummings et al.(2008)]{2008GCN..7462....1C} Cummings, J., Barthelmy, S.~D., Fenimore, E., et al.\ 2008, GRB Coordinates Network, Circular Service, No. 7462, \#1 (2008), 7462
\bibitem[Curran et al.(2007)]{2007A&A...467.1049C} Curran, P.~A., van der Horst, A.~J., Beardmore, A.~P., et al.\ 2007, \aap, 467, 1049
\bibitem[Dai \& Wu(2003)]{2003ApJ...591L..21D} Dai, Z.~G. \& Wu, X.~F.\ 2003, \apjl, 591, L21
\bibitem[Dai \& Lu(1998)]{1998PhRvL..81.4301D} Dai, Z.~G. \& Lu, T.\ 1998a, \prl, 81, 4301
\bibitem[Dai \& Lu(1998)]{1998A&A...333L..87D} Dai, Z.~G. \& Lu, T.\ 1998b, \aap, 333, L87
\bibitem[de Ugarte Postigo et al.(2018)]{2018A&A...620A.190D} de Ugarte Postigo, A., Th{\"o}ne, C.~C., Bensch, K., et al.\ 2018, \aap, 620, A190
\bibitem[Di Matteo et al.(2002)]{2002ApJ...579..706D} Di Matteo, T., Perna, R., \& Narayan, R.\ 2002, \apj, 579, 706
\bibitem[Eichler et al.(1989)]{1989Natur.340..126E} Eichler, D., Livio, M., Piran, T., et al.\ 1989, \nat, 340, 126
\bibitem[Falcone et al.(2007)]{2007ApJ...671.1921F} Falcone, A.~D., Morris, D., Racusin, J., et al.\ 2007, \apj, 671, 1921
\bibitem[Falcone et al.(2006)]{2006ApJ...641.1010F} Falcone, A.~D., Burrows, D.~N., Lazzati, D., et al.\ 2006, \apj, 641, 1010
\bibitem[Fan \& Wei(2011)]{2011ApJ...739...47F} Fan, Y.-Z. \& Wei, D.-M.\ 2011, \apj, 739, 47
\bibitem[Fenimore et al.(2005)]{2005GCN..4217....1F} Fenimore, E., Angelini, L., Barbier, L., et al.\ 2005, GRB Coordinates Network, Circular Service, No. 4217, \#1 (2005), 4217
\bibitem[Fenimore et al.(2006)]{2006GCN..5831....1F} Fenimore, E., Barbier, L., Barthelmy, S.~D., et al.\ 2006, GRB Coordinates Network, Circular Service, No. 5831, \#1 (2006), 5831
\bibitem[Fenimore et al.(2008)]{2008GCN..8297....1F} Fenimore, E., Barthelmy, S.~D., Baumgartner, W., et al.\ 2008, GRB Coordinates Network, Circular Service, No. 8297, \#1 (2008), 8297
\bibitem[Ferrero et al.(2009)]{2009A&A...497..729F} Ferrero, P., Klose, S., Kann, D.~A., et al.\ 2009, \aap, 497, 729
\bibitem[Frail et al.(2001)]{2001ApJ...562L..55F} Frail, D.~A., Kulkarni, S.~R., Sari, R., et al.\ 2001, \apjl, 562, L55
\bibitem[Gao et al.(2015)]{2015ApJ...810..160G} Gao, H., Wang, X.-G., M{\'e}sz{\'a}ros, P., et al.\ 2015, \apj, 810, 160
\bibitem[Gendre et al.(2013)]{2013ApJ...766...30G} Gendre, B., Stratta, G., Atteia, J.~L., et al.\ 2013, \apj, 766, 30
\bibitem[Gorbovskoy et al.(2012)]{2012yCat..74211874G} Gorbovskoy, E.~S., Lipunova, G.~V., Lipunov, V.~M., et al.\ 2012, VizieR Online Data Catalog, J/MNRAS/421/1874
\bibitem[Gorosabel et al.(2005)]{2005NCimC..28..677G} Gorosabel, J., Jel{\'\i}nek, M., de Ugarte Postigo, A., et al.\ 2005, Nuovo Cimento C Geophysics Space Physics C, 28, 677
\bibitem[Gou et al.(2007)]{2007ApJ...668.1083G} Gou, L.-J., Fox, D.~B., \& M{\'e}sz{\'a}ros, P.\ 2007, \apj, 668, 1083
\bibitem[Greiner et al.(2015)]{2015Nature...523...189G} Greiner, J., Mazzali, P. A., Kann, D. A., et al. 2015, Nature, 523, 189
\bibitem[Grupe et al.(2007)]{2007ApJ...662..443G} Grupe, D., Gronwall, C., Wang, X.-Y., et al.\ 2007, \apj, 662, 443.
\bibitem[Gu et al.(2006)]{2006ApJ...643L..87G} Gu, W.-M., Liu, T., \& Lu, J.-F.\ 2006, \apjl, 643, L87
\bibitem[Gu (2015)]{2015ApJ...799...71G} Gu, W.-M. 2015, \apj, 799, 71
\bibitem[Guidorzi et al.(2005)]{2005ApJ...630L.121G} Guidorzi, C., Monfardini, A., Gomboc, A., et al.\ 2005, \apjl, 630, L121
\bibitem[Hartoog et al.(2013)]{2013MNRAS.430.2739H} Hartoog, O.~E., Wiersema, K., Vreeswijk, P.~M., et al.\ 2013, \mnras, 430, 2739
\bibitem[Holland et al.(2006)]{2006GCN..4570....1H} Holland, S.~T., Barthelmy, S., Burrows, D.~N., et al.\ 2006, GRB Coordinates Network, Circular Service, No. 4570, \#1 (2006), 4570
\bibitem[Holland et al.(2002)]{2002AJ....124..639H} Holland, S.~T., Soszy{\'n}ski, I., Gladders, M.~D., et al.\ 2002, \aj, 124, 639
\bibitem[Holland et al.(2010)]{2010ApJ...717..223H} Holland, S.~T., Sbarufatti, B., Shen, R., et al.\ 2010, \apj, 717, 223
\bibitem[Hu et al.(2019)]{2019A&A...632A.100H} Hu, Y.-D., Oates, S.~R., Lipunov, V.~M., et al.\ 2019, \aap, 632, A100
\bibitem[Huang et al.(2017)]{2017PASJ...69...20H} Huang, K., Urata, Y., Takahashi, S., et al.\ 2017, \pasj, 69, 20
\bibitem[Hurkett et al.(2006)]{2006MNRAS.368.1101H} Hurkett, C.~P., Osborne, J.~P., Page, K.~L., et al.\ 2006, \mnras, 368, 1101
\bibitem[Jakobsson et al.(2004)]{2004A&A...427..785J} Jakobsson, P., Hjorth, J., Fynbo, J.~P.~U., et al.\ 2004, \aap, 427, 785
\bibitem[Japelj et al.(2014)]{2014ApJ...785...84J} Japelj, J., Kopa{\v{c}}, D., Kobayashi, S., et al.\ 2014, \apj, 785, 84
\bibitem[Jaunsen et al.(2008)]{2008ApJ...681..453J} Jaunsen, A.~O., Rol, E., Watson, D.~J., et al.\ 2008, \apj, 681, 453
\bibitem[Jin et al.(2013)]{2013ApJ...774..114J} Jin, Z.-P., Covino, S., Della Valle, M., et al.\ 2013, \apj, 774, 114
\bibitem[Kawanaka \& Mineshige(2007)]{2007ApJ...662.1156K} Kawanaka, N. \& Mineshige, S.\ 2007, \apj, 662, 1156
\bibitem[Kawanaka et al.(2013)]{2013ApJ...766...31K} Kawanaka, N., Piran, T., \& Krolik, J.~H.\ 2013, \apj, 766, 31
\bibitem[Kong et al.(2009)]{2009ScChG..52.2047K} Kong, S., Huang, Y., Cheng, K., et al.\ 2009, Science in China: Physics, Mechanics and Astronomy, 52, 2047
\bibitem[Krimm et al.(2005)]{2005GCN..3105....1K} Krimm, H., Barthelmy, S., Barbier, L., et al.\ 2005a, GRB Coordinates Network, Circular Service, No. 3105, \#1 (2005), 3105
\bibitem[Krimm et al.(2005)]{2005GCN..3119....1K} Krimm, H., Sakamoto, T., Barthelmy, S., et al.\ 2005b, GRB Coordinates Network, Circular Service, No. 3119, \#1 (2005), 3119
\bibitem[Krimm et al.(2005)]{2005GCN..3134....1K} Krimm, H., Barthelmy, S., Barbier, L., et al.\ 2005c, GRB Coordinates Network, Circular Service, No. 3134, \#1 (2005), 3134
\bibitem[Krimm et al.(2005)]{2005GCN..4020....1K} Krimm, H., Barbier, L., Barthelmy, S., et al.\ 2005d, GRB Coordinates Network, Circular Service, No. 4020, \#1 (2005), 4020
\bibitem[Krimm et al.(2005)]{2005GCN..4260....1K} Krimm, H., Ajello, M., Barbier, L., et al.\ 2005e, GRB Coordinates Network, Circular Service, No. 4260, \#1 (2005), 4260
\bibitem[Krimm et al.(2006)]{2006GCN..5860....1K} Krimm, H., Barbier, L., Barthelmy, S.~D., et al.\ 2006, GRB Coordinates Network, Circular Service, No. 5860, \#1 (2006), 5860
\bibitem[Krimm et al.(2007a)]{2007ApJ...665..554K} Krimm, H.~A., Granot, J., Marshall, F.~E., et al.\ 2007a, \apj, 665, 554
\bibitem[Krimm et al.(2007b)]{2007GCN..6868....1K} Krimm, H., Barthelmy, S.~D., Cummings, J., et al.\ 2007b, GRB Coordinates Network, Circular Service, No. 6868, \#1 (2007), 6868
\bibitem[Krimm et al.(2010)]{2010GCN.11094....1K} Krimm, H.~A., Barthelmy, S.~D., Baumgartner, W.~H., et al.\ 2010, GRB Coordinates Network, Circular Service, No. 11094, \#1 (2010), 11094
\bibitem[Krimm et al.(2019)]{2019GCN.23724....1K} Krimm, H.~A., Barthelmy, S.~D., Cummings, J.~R., et al.\ 2019, GRB Coordinates Network, Circular Service, No. 23724, \#1 (2019), 23724
\bibitem[Krimm et al.(2017)]{2017GCN.21112....1K} Krimm, H.~A., Barthelmy, S.~D., Baumgartner, W.~H., et al.\ 2017, GRB Coordinates Network, Circular Service, No. 21112, \#1 (2017), 21112
\bibitem[Krimm et al.(2014)]{2014GCN.15962....1K} Krimm, H.~A., Barthelmy, S.~D., Baumgartner, W.~H., et al.\ 2014a, GRB Coordinates Network, Circular Service, No. 15962, \#1 (2014), 15962
\bibitem[Krimm et al.(2014)]{2014GCN.17083....1K} Krimm, H.~A., Barthelmy, S.~D., Baumgartner, W.~H., et al.\ 2014b, GRB Coordinates Network, Circular Service, No. 17083, \#1 (2014), 17083
\bibitem[Kumar \& Zhang(2015)]{2015PhR...561....1K} Kumar, P. \& Zhang, B.\ 2015, \physrep, 561, 1
\bibitem[Lee et al.(2000)]{2000PhR...325...83L} Lee, H.~K., Wijers, R.~A.~M.~J., \& Brown, G.~E.\ 2000a, \physrep, 325, 83
\bibitem[Lee et al.(2000)]{2000ApJ...536..416L} Lee, H.~K., Brown, G.~E., \& Wijers, R.~A.~M.~J.\ 2000b, \apj, 536, 416
\bibitem[Lei et al.(2009)]{} Lei, W.-H., Wang, D.-X., Zhang, L., et al. 2009, \apj, 700, 1970
\bibitem[Lei \& Zhang(2011)]{2011ApJ...740L..27L} Lei, W.-H. \& Zhang, B.\ 2011, \apjl, 740, L27
\bibitem[Lei et al.(2013)]{2013ApJ...765..125L} Lei, W.-H., Zhang, B., \& Liang, E.-W.\ 2013, \apj, 765, 125
\bibitem[Lei et al.(2017)]{2017ApJ...849...47L} Lei, W.-H., Zhang, B., Wu, X.-F., et al.\ 2017, \apj, 849, 47
\bibitem[Li et al.(2018)]{2018ApJS..236...26L} Li, L., Wu, X.-F., Lei, W.-H., et al.\ 2018, \apjs, 236, 26
\bibitem[Liang et al.(2006)]{2006ApJ...646..351L} Liang, E.~W., Zhang, B., O'Brien, P.~T., et al.\ 2006, \apj, 646, 351
\bibitem[Liang et al.(2013)]{2013ApJ...774...13L} Liang, E.-W., Li, L., Gao, H., et al.\ 2013, \apj, 774, 13
\bibitem[Liang et al.(2010)]{2010ApJ...725.2209L} Liang, E.-W., Yi, S.-X., Zhang, J., et al.\ 2010, \apj, 725, 2209
\bibitem[Liang et al.(2008)]{2008ApJ...675..528L} Liang, E.-W., Racusin, J.~L., Zhang, B., et al.\ 2008, \apj, 675, 528
\bibitem[Lien et al.(2018)]{2018GCN.23321....1L} Lien, A.~Y., Barthelmy, S.~D., Cummings, J.~R., et al.\ 2018, GRB Coordinates Network, Circular Service, No. 23321, \#1 (2018/October-0), 23321
\bibitem[Liu et al.(2007)]{2007ApJ...661.1025L} Liu, T., Gu, W.-M., Xue, L., et al.\ 2007, \apj, 661, 1025
\bibitem[Liu et al.(2015)]{2015ApJ...806...58L} Liu, T., Lin, Y.-Q., Hou, S.-J., et al.\ 2015, \apj, 806, 58
\bibitem[Liu et al.(2018)]{2018ApJ...852...20L} Liu, T., Song, C.-Y., Zhang, B., et al.\ 2018, \apj, 852, 20
\bibitem[Liu et al.(2016)]{2016ApJ...821..132L} Liu, T., Xue, L., Zhao, X.-H., et al.\ 2016, \apj, 821, 132
\bibitem[Liu et al.(2017)]{2017NewAR..79....1L} Liu, T., Gu, W.-M., \& Zhang, B.\ 2017, \nar, 79, 1
\bibitem[Liu et al.(2015)]{2015ApJS..218...12L} Liu, T., Hou, S.-J., Xue, L., et al.\ 2015, \apjs, 218, 12
\bibitem[Liu et al.(2008)]{2008A&A...487..503L} Liu, X.~W., Wu, X.~F., \& Lu, T.\ 2008, \aap, 487, 503
\bibitem[Lu et al.(2012)]{2012ApJ...745..168L} Lu, R.-J., Wei, J.-J., Qin, S.-F., et al.\ 2012, \apj, 745, 168
\bibitem[L{\"u} \& Zhang(2014)]{2014ApJ...785...74L} L{\"u}, H.-J. \& Zhang, B.\ 2014, \apj, 785, 74
\bibitem[L{\"u} et al.(2015)]{2015ApJ...805...89L} L{\"u}, H.-J., Zhang, B., Lei, W.-H., et al.\ 2015, \apj, 805, 89
\bibitem[MacFadyen \& Woosley(1999)]{1999ApJ...524..262M} MacFadyen, A.~I. \& Woosley, S.~E.\ 1999, \apj, 524, 262
\bibitem[Maiorano et al.(2005)]{2005A&A...438..821M} Maiorano, E., Masetti, N., Palazzi, E., et al.\ 2005, \aap, 438, 821
\bibitem[Markwardt et al.(2008)]{2008GCN..7549....1M} Markwardt, C., Barthelmy, S.~D., Cummings, J., et al.\ 2008, GRB Coordinates Network, Circular Service, No. 7549, \#1 (2008), 7549
\bibitem[Markwardt et al.(2006)]{2006GCN..5174....1M} Markwardt, C., Barbier, L., Barthelmy, S., et al.\ 2006, GRB Coordinates Network, Circular Service, No. 5174, \#1 (2006), 5174
\bibitem[Markwardt et al.(2007)]{2007GCN..6274....1M} Markwardt, C., Barbier, L., Barthelmy, S.~D., et al.\ 2007a, GRB Coordinates Network, Circular Service, No. 6274, \#1 (2007), 6274
\bibitem[Markwardt et al.(2007)]{2007GCN..6748....1M} Markwardt, C., Barbier, L., Barthelmy, S.~D., et al.\ 2007b, GRB Coordinates Network, Circular Service, No. 6748, \#1 (2007), 6748
\bibitem[Markwardt et al.(2007)]{2007GCN..6877....1M} Markwardt, C., Barthelmy, S.~D., Cummings, J., et al.\ 2007c, GRB Coordinates Network, Circular Service, No. 6877, \#1 (2007), 6877
\bibitem[\protect\citeauthoryear{McKinney}{2005}]{2005ApJ...630L...5M} McKinney J.~C., 2005, ApJL, 630, L5
\bibitem[Melandri et al.(2014)]{2014A&A...572A..55M} Melandri, A., Virgili, F.~J., Guidorzi, C., et al.\ 2014, \aap, 572, A55
\bibitem[Metzger(2010)]{2010MNRAS.409..284M} Metzger, B.~D.\ 2010, \mnras, 409, 284
\bibitem[Miller \& Miller(2015)]{2015PhR...548...1M} Miller, M.~C. \& Miller, J.~M. 2015, PhR, 548, 1
\bibitem[Minaev \& Pozanenko(2020)]{2020MNRAS.492.1919M} Minaev, P.~Y. \& Pozanenko, A.~S.\ 2020, \mnras, 492, 1919
\bibitem[Mirabal et al.(2007)]{2007ApJ...661L.127M} Mirabal, N., Halpern, J.~P., \& O'Brien, P.~T.\ 2007, \apjl, 661, L127
\bibitem[M{\'e}sz{\'a}ros \& Rees(1997)]{1997ApJ...476..232M} M{\'e}sz{\'a}ros, P. \& Rees, M.~J.\ 1997, \apj, 476, 232
\bibitem[Nagataki(2011)]{2011PASJ...63.1243N} Nagataki, S.\ 2011, \pasj, 63, 1243
\bibitem[Narayan et al.(2001)]{2001ApJ...557..949N} Narayan, R., Piran, T., \& Kumar, P.\ 2001, \apj, 557, 949
\bibitem[Narayan et al.(1992)]{1992ApJ...395L..83N} Narayan, R., Paczynski, B., \& Piran, T.\ 1992, \apjl, 395, L83
\bibitem[Nousek et al.(2006)]{2006ApJ...642..389N} Nousek, J.~A., Kouveliotou, C., Grupe, D., et al.\ 2006, \apj, 642, 389
\bibitem[Paczynski(1986)]{1986ApJ...308L..43P} Paczynski, B.\ 1986, \apjl, 308, L43
\bibitem[Palmer et al.(2007)]{2007GCN..6643....1P} Palmer, D., Barbier, L., Barthelmy, S.~D., et al.\ 2007, GRB Coordinates Network, Circular Service, No. 6643, \#1 (2007), 6643
\bibitem[Palmer et al.(2005)]{2005GCN..3737....1P} Palmer, D., Barbier, L., Barthelmy, S., et al.\ 2005, GRB Coordinates Network, Circular Service, No. 3737, \#1 (2005), 3737
\bibitem[Palmer et al.(2017)]{2017GCN.20999....1P} Palmer, D.~M., Barthelmy, S.~D., Cummings, J.~R., et al.\ 2017, GRB Coordinates Network, Circular Service, No. 20999, \#1 (2017), 20999
\bibitem[Palmer et al.(2008)]{2008GCN..8351....1P} Palmer, D.~M., Barthelmy, S.~D., Baumgartner, W.~H., et al.\ 2008, GRB Coordinates Network, Circular Service, No. 8351, \#1 (2008), 8351
\bibitem[Palmer et al.(2015)]{2015GCN.18496....1P} Palmer, D.~M., Barthelmy, S.~D., Cummings, J.~R., et al.\ 2015, GRB Coordinates Network, Circular Service, No. 18496, \#1 (2015), 18496
\bibitem[Palmer et al.(2012)]{2012GCN.13536....1P} Palmer, D.~M., Barthelmy, S.~D., Baumgartner, W.~H., et al.\ 2012, GRB Coordinates Network, Circular Service, No. 13536, \#1 (2012), 13536
\bibitem[Palmer et al.(2016)]{2016GCN.20308....1P} Palmer, D.~M., Barthelmy, S.~D., Cummings, J.~R., et al.\ 2016, GRB Coordinates Network, Circular Service, No. 20308, \#1 (2016), 20308
\bibitem[Parsons et al.(2006)]{2006GCN..5370....1P} Parsons, A., Barbier, L., Barthelmy, S.~D., et al.\ 2006, GRB Coordinates Network, Circular Service, No. 5370, \#1 (2006), 5370
\bibitem[Perri et al.(2007)]{2007A&A...471...83P} Perri, M., Guetta, D., Antonelli, L.~A., et al.\ 2007, \aap, 471, 83
\bibitem[Piran(2004)]{2004RvMP...76.1143P} Piran, T.\ 2004, Reviews of Modern Physics, 76, 1143
\bibitem[Piro et al.(2005)]{2005ApJ...623..314P} Piro, L., De Pasquale, M., Soffitta, P., et al.\ 2005, \apj, 623, 314
\bibitem[Popham et al.(1999)]{1999ApJ...518..356P} Popham, R., Woosley, S.~E., \& Fryer, C.\ 1999, \apj, 518, 356
\bibitem[Racusin et al.(2009)]{2009ApJ...698...43R} Racusin, J.~L., Liang, E.~W., Burrows, D.~N., et al.\ 2009, \apj, 698, 43
\bibitem[Rhoads(1999)]{1999ApJ...525..737R} Rhoads, J.~E.\ 1999, \apj, 525, 737
\bibitem[Rowlinson et al.(2014)]{2014MNRAS.443.1779R} Rowlinson, A., Gompertz, B.~P., Dainotti, M., et al.\ 2014, \mnras, 443, 1779
\bibitem[Rowlinson et al.(2013)]{2013MNRAS.430.1061R} Rowlinson, A., O'Brien, P.~T., Metzger, B.~D., et al.\ 2013, \mnras, 430, 1061
\bibitem[Ruffini et al.(2018)]{2018ApJ...852...53R} Ruffini, R., Wang, Y., Aimuratov, Y., et al.\ 2018, \apj, 852, 53
\bibitem[Sagar et al.(2000)]{2000BASI...28...15S} Sagar, R., Mohan, V., Pandey, A.~K., et al.\ 2000, Bulletin of the Astronomical Society of India, 28, 15
\bibitem[Sakamoto et al.(2005)]{2005GCN..3173....1S} Sakamoto, T., Barthelmy, S., Barbier, L., et al.\ 2005a, GRB Coordinates Network, Circular Service, No. 3173, \#1 (2005), 3173
\bibitem[Sakamoto et al.(2005)]{2005GCN..3730....1S} Sakamoto, T., Markwardt, C., Barbier, L., et al.\ 2005b, GRB Coordinates Network, Circular Service, No. 3730, \#1 (2005), 3730
\bibitem[Sakamoto et al.(2006)]{2006GCN..5534....1S} Sakamoto, T., Barbier, L., Barthelmy, S.~D., et al.\ 2006, GRB Coordinates Network, Circular Service, No. 5534, \#1 (2006), 5534
\bibitem[Sakamoto et al.(2008)]{2008GCN..8082....1S} Sakamoto, T., Barthelmy, S.~D., Baumgartner, W., et al.\ 2008, GRB Coordinates Network, Circular Service, No. 8082, \#1 (2008), 8082
\bibitem[Sakamoto et al.(2018)]{2018GCN.23357....1S} Sakamoto, T., Barthelmy, S.~D., Cummings, J.~R., et al.\ 2018, GRB Coordinates Network, Circular Service, No. 23357, \#1 (2018), 23357
\bibitem[Sakamoto et al.(2016)]{2016GCN.19276....1S} Sakamoto, T., Barthelmy, S.~D., Cummings, J.~R., et al.\ 2016b, GRB Coordinates Network, Circular Service, No. 19276, \#1 (2016), 19276
\bibitem[Sakamoto et al.(2016)]{2016GCN.19106....1S} Sakamoto, T., Barthelmy, S.~D., Cummings, J.~R., et al.\ 2016a, GRB Coordinates Network, Circular Service, No. 19106, \#1 (2016), 19106
\bibitem[Sakamoto et al.(2009)]{2009GCN..8986....1S} Sakamoto, T., Barthelmy, S.~D., Baumgartner, W.~H., et al.\ 2009, GRB Coordinates Network, Circular Service, No. 8986, \#1 (2009), 8986
\bibitem[Sakamoto et al.(2014)]{2014GCN.16258....1S} Sakamoto, T., Barthelmy, S.~D., Baumgartner, W.~H., et al.\ 2014, GRB Coordinates Network, Circular Service, No. 16258, \#1 (2014), 16258
\bibitem[Samuelsson et al.(2020)]{2020arXiv200502417S} Samuelsson, F., B{\'e}gu{\'e}, D., Ryde, F., et al.\ 2020, arXiv:2005.02417
\bibitem[Sari et al.(1998)]{1998ApJ...497L..17S} Sari, R., Piran, T., \& Narayan, R.\ 1998, \apjl, 497, L17
\bibitem[Sari et al.(1999)]{1999ApJ...519L..17S} Sari, R., Piran, T., \& Halpern, J.~P.\ 1999, \apjl, 519, L17
\bibitem[Sato et al.(2005)]{2005PASJ...57.1031S} Sato, R., Sakamoto, T., Kataoka, J., et al.\ 2005, \pasj, 57, 1031
\bibitem[Si et al.(2018)]{2018ApJ...863...50S} Si, S.-K., Qi, Y.-Q., Xue, F.-X., et al.\ 2018, \apj, 863, 50
\bibitem[Soderberg et al.(2007)]{2007ApJ...661..982S} Soderberg, A.~M., Nakar, E., Cenko, S.~B., et al.\ 2007, \apj, 661, 982
\bibitem[Song et al.(2015)]{2015ApJ...815...54S}Song C.-Y., Liu T., Gu W.-M., et al.\ 2015,\apj, 815, 54
\bibitem[Song et al.(2016)]{2016MNRAS.458.1921S} Song, C.-Y., Liu, T., Gu, W.-M., et al.\ 2016, \mnras, 458, 1921
\bibitem[Song et al.(2018)]{2018MNRAS.477.2173S} Song, C.-Y., Liu, T., \& Li, A.\ 2018, \mnras, 477, 2173
\bibitem[Song, Liu, \& Wei (2020)]{2020MNRAS.494.3962S} Song C.-Y., Liu T., Wei Y.-F.\ 2020, \mnras, 494, 3962.
\bibitem[Stamatikos et al.(2007)]{2007GCN..7029....1S} Stamatikos, M., Barthelmy, S.~D., Cummings, J., et al.\ 2007b, GRB Coordinates Network, Circular Service, No. 7029, \#1 (2007), 7029
\bibitem[Stamatikos et al.(2007)]{2007GCN..6326....1S} Stamatikos, M., Barbier, L., Barthelmy, S.~D., et al.\ 2007a, GRB Coordinates Network, Circular Service, No. 6326, \#1 (2007), 6326
\bibitem[Stamatikos et al.(2006)]{2006GCN..5289....1S} Stamatikos, M., Barbier, L., Barthelmy, S., et al.\ 2006a, GRB Coordinates Network, Circular Service, No. 5289, \#1 (2006), 5289
\bibitem[Stamatikos et al.(2006)]{2006GCN..5459....1S} Stamatikos, M., Barbier, L., Barthelmy, S.~D., et al.\ 2006b, GRB Coordinates Network, Circular Service, No. 5459, \#1 (2006), 5459
\bibitem[Stamatikos et al.(2006)]{2006GCN..5639....1S} Stamatikos, M., Barbier, L., Barthelmy, S.~D., et al.\ 2006c, GRB Coordinates Network, Circular Service, No. 5639, \#1 (2006), 5639
\bibitem[Stamatikos et al.(2012)]{2012GCN.12884....1S} Stamatikos, M., Barthelmy, S.~D., Baumgartner, W.~H., et al.\ 2012, GRB Coordinates Network, Circular Service, No. 12884, \#1 (2012), 12884

\bibitem[Stamatikos et al.(2009)]{2009GCN.10197....1S} Stamatikos, M., Barthelmy, S.~D., Baumgartner, W.~H., et al.\ 2009, GRB Coordinates Network, Circular Service, No. 10197, \#1 (2009), 10197

\bibitem[Stratta et al.(2005)]{2005A&A...441...83S} Stratta, G., Perna, R., Lazzati, D., et al.\ 2005, \aap, 441, 83
\bibitem[Swenson et al.(2010)]{2010ApJ...718L..14S} Swenson, C.~A., Maxham, A., Roming, P.~W.~A., et al.\ 2010, \apjl, 718, L14
\bibitem[Tchekhovskoy et al.(2011)]{2011MNRAS.418L..79T} Tchekhovskoy, A., Narayan, R., \& McKinney, J.~C.\ 2011, \mnras, 418, L79
\bibitem[Troja et al.(2015)]{2015ApJ...803...10T} Troja, E., Piro, L., Vasileiou, V., et al.\ 2015, \apj, 803, 10
\bibitem[Tueller et al.(2005)]{2005GCN..3803....1T} Tueller, J., Markwardt, C., Barbier, L., et al.\ 2005, GRB Coordinates Network, Circular Service, No. 3803, \#1 (2005), 3803
\bibitem[Tueller et al.(2008)]{2008GCN..7402....1T} Tueller, J., Barthelmy, S.~D., Cummings, J., et al.\ 2008a, GRB Coordinates Network, Circular Service, No. 7402, \#1 (2008), 7402
\bibitem[Tueller et al.(2008)]{2008GCN..7604....1T} Tueller, J., Barthelmy, S.~D., Baumgartner, W., et al.\ 2008b, GRB Coordinates Network, Circular Service, No. 7604, \#1 (2008), 7604

\bibitem[Tueller et al.(2008)]{2008GCN..7969....1T} Tueller, J., Barthelmy, S.~D., Baumgartner, W., et al.\ 2008c, GRB Coordinates Network, Circular Service, No. 7969, \#1 (2008), 7969
\bibitem[Ukwatta et al.(2007)]{2007GCN..6842....1U} Ukwatta, T., Barbier, L., Barthelmy, S.~D., et al.\ 2007, GRB Coordinates Network, Circular Service, No. 6842, \#1 (2007), 6842

\bibitem[Ukwatta et al.(2012)]{2012GCN.13220....1U} Ukwatta, T.~N., Barthelmy, S.~D., Baumgartner, W.~H., et al.\ 2012, GRB Coordinates Network, Circular Service, No. 13220, \#1 (2012), 13220
\bibitem[Ukwatta et al.(2008)]{2008GCN..7289....1U} Ukwatta, T., Barthelmy, S.~D., Cummings, J., et al.\ 2008a, GRB Coordinates Network, Circular Service, No. 7289, \#1 (2008), 7289
\bibitem[Ukwatta et al.(2008)]{2008GCN..8599....1U} Ukwatta, T.~N., Barthelmy, S.~D., Baumgartner, W.~H., et al.\ 2008b, GRB Coordinates Network, Circular Service, No. 8599, \#1 (2008), 8599
\bibitem[Usov(1992)]{1992Natur.357..472U} Usov, V.~V.\ 1992, \nat, 357, 472

\bibitem[Wang et al.(2003)]{2003A&A...401..593W} Wang, X.~Y., Dai, Z.~G., \& Lu, T.\ 2003, \aap, 401, 593
\bibitem[Wang et al.(2018)]{2018ApJ...859..160W} Wang, X.-G., Zhang, B., Liang, E.-W., et al.\ 2018, \apj, 859, 160
\bibitem[Wijers et al.(1999)]{1999ApJ...523L..33W} Wijers, R.~A.~M.~J., Vreeswijk, P.~M., Galama, T.~J., et al.\ 1999, \apjl, 523, L33
\bibitem[Woosley(1993)]{1993ApJ...405..273W} Woosley, S.~E.\ 1993, \apj, 405, 273
\bibitem[Wu et al.(2004)]{2004ApJ...615..359W} Wu, X.~F., Dai, Z.~G., \& Liang, E.~W.\ 2004, \apj, 615, 359
\bibitem[Wu et al.(2003)]{2003MNRAS.342.1131W} Wu, X.~F., Dai, Z.~G., Huang, Y.~F., et al.\ 2003, \mnras, 342, 1131
\bibitem[Wu et al.(2013)]{2013ApJ...767L..36W} Wu, X.-F., Hou, S.-J., \& Lei, W.-H.\ 2013, \apjl, 767, L36
\bibitem[Xi et al.(2017)]{2017RAA....17...53X} Xi, S.-Q., Yi, S.-X., Zou, Y.-C., et al.\ 2017, Research in Astronomy and Astrophysics, 17, 053
\bibitem[Xie et al.(2017)]{2017ApJ...838..143X} Xie, W., Lei, W.-H., \& Wang, D.-X.\ 2017, \apj, 838, 143
\bibitem[Xie et al.(2016)]{2016ApJ...833..129X} Xie, W., Lei, W.-H., \& Wang, D.-X.\ 2016, \apj, 833, 129
\bibitem[Xue et al.(2013)]{2013ApJS..207...23X} Xue, L., Liu, T., Gu, W.-M., et al.\ 2013, \apjs, 207, 23
\bibitem[Xue et al.(2019)]{2019ApJ...876...77X} Xue, L., Zhang, F.-W., \& Zhu, S.-Y.\ 2019, \apj, 876, 77
\bibitem[Yi et al.(2017)]{2017JHEAp..13....1Y} Yi, S.-X., Lei, W.-H., Zhang, B., et al.\ 2017b, Journal of High Energy Astrophysics, 13, 1
\bibitem[Yi et al.(2013)]{2013ApJ...776..120Y} Yi, S.-X., Wu, X.-F., \& Dai, Z.-G.\ 2013, \apj, 776, 120
\bibitem[Yi et al.(2017)]{2017ApJ...844...79Y} Yi, S.-X., Yu, H., Wang, F.~Y., et al.\ 2017a, \apj, 844, 79
\bibitem[Yi et al.(2016)]{2016ApJS..224...20Y} Yi, S.-X., Xi, S.-Q., Yu, H., et al.\ 2016, \apjs, 224, 20
\bibitem[Yi et al.(2015)]{2015ApJ...807...92Y} Yi, S.-X., Wu, X.-F., Wang, F.-Y., et al.\ 2015, \apj, 807, 92
\bibitem[Yi et al.(2020)]{2020ApJ...895...94Y} Yi, S.-X., Wu, X.-F., Zou, Y.-C., et al.\ 2020, \apj, 895, 94
\bibitem[Zalamea \& Beloborodov(2011)]{2011MNRAS.410.2302Z} Zalamea, I. \& Beloborodov, A.~M.\ 2011, \mnras, 410, 2302
\bibitem[Zhang(2018)]{2018pgrb.book.....Z} Zhang, B.\ 2018, The Physics of Gamma-Ray Bursts by Bing Zhang. ISBN: 978-1-139-22653-0. Cambridge Univeristy Press, 2018.
\bibitem[Zhang et al.(2006)]{2006ApJ...642..354Z} Zhang, B., Fan, Y.~Z., Dyks, J., et al.\ 2006, \apj, 642, 354
\bibitem[Zhang(2007)]{2007ChJAA...7....1Z} Zhang, B.\ 2007, \cjaa, 7, 1
\bibitem[Zhang \& M{\'e}sz{\'a}ros(2001)]{2001ApJ...552L..35Z} Zhang, B. \& M{\'e}sz{\'a}ros, P.\ 2001, \apjl, 552, L35
\bibitem[Zhang et al.(2003)]{2003ApJ...586...356Z}Zhang, W., Woosley, S. E., \& MacFadyen, A. I. 2003, \apj, 586, 356

\bibitem[Zheng et al.(2012)]{2012ApJ...751...90Z} Zheng, W., Shen, R.~F., Sakamoto, T., et al.\ 2012, \apj, 751, 90
\bibitem[Zhao et al.(2020)]{2020ApJ...900..112Y} Zhao, W., Zhang, J.~C., Zhang, Q.~X., et al.\ 2020, \apj, 900, 112
\bibitem[Zhou et al.(2020)]{2020IJMPD..2950043Z} Zhou, Q.-Q., Yi, S.-X., Huang, X.-L., et al.\ 2020, International Journal of Modern Physics D, 29, 2050043
\bibitem[Zou et al.(2005)]{2005MNRAS.363...93Z} Zou, Y.~C., Wu, X.~F., \& Dai, Z.~G.\ 2005, \mnras, 363, 93


\end{thebibliography}
\end{document}